# Frequency- adaptive control of a three-phase single-stage grid-connected photovoltaic system under grid voltage sags

**Alexis B. Rey-Boué[1*], N. F. Guerrero-Rodríguez[2], Johannes Stöckl[3], and Thomas I. Strasser[3]**

[1]  Department of Electronics, Computers Technology and Projects, Universidad Politécnica de Cartagena, c/Dr. Fleming, s/n, 30202 Cartagena, Murcia, Spain. Tel.: +34 968 325928; fax: +34 968 326400; alexis.rey@upct.es

[2]  Engineering Sciences, Pontificia Universidad Católica Madre y Maestra PUCMM, Av. Abraham Lincoln Esq. Romulo Betancourt, 2748 Santo Domingo, Dominican Republic. Tel.: +1 (809) 535-0111 ext.2314; fax: +1 (809) 534-7060; nf.guerrero@ce.pucmm.edu.do

[3]  AIT Austrian Institute of Technology, Center for Energy – Electric Energy Systems, Giefinggasse 2, 1210 Vienna, Austria. Tel.: +43(0) 50550-6032; fax: +43(0) 50550-6390; johannes.stoeckl@ait.ac.at

[3]  AIT Austrian Institute of Technology, Center for Energy – Electric Energy Systems, Giefinggasse 2, 1210 Vienna, Austria. Tel.: +43(0) 50550-6279; fax: +43(0) 50550-6390; thomas.i.strasser@ieee.org

*  Correspondence author: alexis.rey@upct.es; tel.: +34 968 325928

**Abstract:** The low-voltage ride-through service is carried out in this paper according to the voltage profile described by the IEC 61400-21 European normative when short-duration voltage sags happen, and some instantaneous reactive power is delivered to the grid in accordance with the Spanish grid code; the mandatory limitation of the amplitude of the three-phase inverter currents to its nominal value is carried out with a novel control strategy, in which a certain amount of instantaneous constant active power can also be delivered to the grid when small or moderate voltage sags happen. A Multiple second order generalized integrator frequency-locked loop synchronization algorithm is employed in order to estimate the system frequency without harmonic distortions, as well as to output the positive- and the negative- sequence of the $\alpha\beta$ quantities of the three-phase grid voltages when balanced and unbalanced voltage sags happen in a frequency- adaptive scheme. The current control is carried out in the stationary reference frame, which guarantees the cancellation of the harmonic distortions in the utility grid currents using a Harmonic compensation structure, and the implementation of a constant active power control in order to protect the DC link capacitor from thermal stresses avoiding the appearance of large harmonic distortions at twice the fundamental frequency in the DC link voltage. A case study of a three-phase single-stage grid-connected PV system with a maximum apparent power about 500 kVA is tested with several simulations using MATLAB/SIMULINK firstly, and secondly, with some experiments using the Controller hardware-in-the-loop (CHIL) simulation technique for several types of voltage sags in order to do the final validation of the control algorithms.

**Keywords:** CHIL simulation technique, constant active power control, frequency- adaptive grid-connected PV system, low-voltage ride-through, MSOGI-FLL.

**Nomenclature**
**Acronyms**

| | |
|---|---|
| CHIL | Controller hardware-in-the-loop |
| CRG | Current References Generator |
| DSC | Delayed signal cancellation |
| DSOGI-FLL | Dual second order generalized integrator FLL |
| DSOGI-QSG-i | Dual second order generalized integrators with quadrature signal generator-harmonic i |
| FLL | Frequency- locked loop |
| HC | Harmonic compensator |
| HDN | Harmonic decoupling network |
| LVRT | Low-voltage ride-through |
| MPPT | Maximum Power Point Tracking |

| MSOGI-FLL | Multiple second order generalized integrator FLL |
| NS | Negative- sequence |
| PCC | Point of common coupling |
| PF | Power factor |
| PI | Proportional integral |
| PLL | Phase-locked loop |
| PNSC-i | Positive- negative- sequence calculator-harmonic i |
| PNSD | Positive- negative- sequence detector |
| PR | Proportional resonant |
| PS | Positive- sequence |
| PV | Photovoltaic |
| PWM | Pulse Width Modulation |
| SOGI | Second order generalized integrator |
| SRF | Synchronous reference frame |
| STC | Standard test conditions |
| VSI | Voltage source inverter |

**Symbols**

| $1- K_P (1- K_Q)$ | Split of the delivered active (reactive) power current due to the NS component of the grid voltages |
| $c(t)$ | Triangular carrier signal for the PWM implementation |
| $C_{link}$ | Link capacitor |
| $f_{CI}$ | Crossover frequency of the inner current loop |
| $f_{CV}$ | Crossover frequency of the outer voltage loop |
| $f_{sw}$ | Switching frequency |
| $\mathbf{i^*}$ | Reference inverter current vector |
| $i^*_{\alpha\beta}\pm$ | $\alpha\beta$ quantities corresponding to the PS and the NS components of the instantaneous three-phase inverter current reference commands |
| $i^*_{\alpha\beta P^*(Q^*)}$ | $\alpha\beta$ quantities corresponding to $P^*$( $Q^*$) of the instantaneous three-phase inverter current reference commands |
| $i_{Clink}$ | DC current through the link capacitor |
| $i_{DC}$ | DC current at the input of the inverter |
| $i_P$ | DC current at the output of the PV generator |
| $\mathbf{i^*_P}$ ($\mathbf{i^*_Q}$) | Reference inverter current vector according to $P^*$ ($Q^*$) |
| $I_{sc}$ | Short circuit current at PCC |
| $i_{\alpha\beta}$ | $\alpha\beta$ quantities of the instantaneous three-phase inverter currents |
| $i_{\alpha\beta}^*$ | $\alpha\beta$ quantities of the instantaneous three-phase inverter current reference commands |
| $K$ | Gain of the normalization block inside the FLL |
| $K/\hat{i}$ | Gain of the DSOGI-QSG-i |
| $K_{IVDC}$ | Integral constant of the DC voltage regulator |
| $K_{I\alpha\beta}$ | Integral constant of the PR controller |
| $K_P (K_Q)$ | Split of the delivered active (reactive) power current due to the PS component of the grid voltages |
| $K_{PVDC}$ | Proportional constant of the DC voltage regulator |
| $K_{PWM}$ | Inverter gain |
| $K_{P\alpha\beta}$ | Proportional constant of the PR controller |
| $L$ | Line inductance of the filter |
| $m_{abc}(t)$ | Three-phase modulating signals for the PWM implementation |
| $m_{\alpha\beta}$ | $\alpha\beta$ quantities of the modulating signals for the PWM implementation |
| $P(Q)$ | Instantaneous active (reactive) power |
| $P^*(Q^*)$ | Active (reactive) power reference command |
| $P_0 (Q_0)$ | Average active (reactive) power |
| $P_{0-} (Q_{0-})$ | Average active (reactive) power for the NS components |
| $P_{0+} (Q_{0+})$ | Average active (reactive) power for the PS components |

| $P_{fault}$ | Instantaneous active power during faulty operation mode |
|---|---|
| $P_{max}$ | Maximum active power at the output of the PV generator for a specific irradiance and temperature |
| $\bar{P}$ ($\bar{Q}$) | Oscillating active (reactive) power at $2\omega_0´$ |
| $PM_I$ | Phase margin of the inner current loop |
| $PM_V$ | Phase margin of the outer voltage loop |
| $S_{fault}$ | Maximum apparent power during faulty operation mode |
| $S_{nom}$ | Maximum apparent power |
| $u^+_{g1(\alpha\beta)}´$ ($u^-_{g1(\alpha\beta)}´$) | Estimated $\alpha\beta$ quantities corresponding to the fundamental PS (NS) component of the instantaneous three-phase grid voltages |
| $u^+_{gn(\alpha\beta)}´$ ($u^-_{gn(\alpha\beta)}´$) | Estimated $\alpha\beta$ quantities corresponding to the $n^{th}$ harmonic PS (NS) component of the instantaneous three-phase grid voltages |
| $u_{g(\alpha\beta)}$ | $\alpha\beta$ quantities corresponding to the instantaneous three-phase grid voltages |
| $u_{g(\alpha\beta)\pm}$ | $\alpha\beta$ quantities corresponding to the instantaneous PS and NS components of the three-phase grid voltages |
| $\boldsymbol{u_{g\alpha\beta}}\perp$ | In-quadrature components of the $\alpha\beta$ quantities of the three-phase grid voltages |
| $u_{gnom}$ | Nominal *root-mean-square (rms)* value of the phase-to-neutral grid voltages |
| $u_{grst}$ | Instantaneous three-phase grid voltages |
| $V*_{DC}$ | DC link voltage reference command |
| $v_{DC}$ | DC link voltage |
| $V_{fault}$ | Depth of the voltage sag |
| $V_{oc}$ | Open circuit voltage at PCC |
| $v_P$ | DC voltage at the output of the PV generator |
| $Z$ | Grid impedance |
| $\Gamma$ | Gain of the FLL |
| $\omega_0$ | Fundamental nominal frequency |
| $\omega_0´$ | Estimated system frequency of the instantaneous three-phase grid voltages |
| $\omega_c$ | Cut-off frequency of the PR controller |

## 1. Introduction

The penetration of renewable sources in the energy mix [1,2], such as wind and photovoltaic (PV) generation, significantly increased in the last decade, where the stability and the reliability of the power system together with the power quality is mandatory, but also a challenging task for the system operators. Therefore, several countries regulate the grid-connection of renewables to the utility grid by its own grid codes [3–6] and the so-called low-voltage ride-through (LVRT) capability arises as an ancillary service to the main control algorithms. The LVRT capability will avoid the disconnection of the grid-connected PV system from the utility grid during very short-duration voltage sags. Such a service is not only mandatory, but also a challenging task and several works have being carried out in order to improve its behavior [7–11]. Its main goals are:

1) To calculate the positive- and the negative- sequence (PS and the NS) components of the three-phase grid voltages and currents.

2) To deliver a certain amount of instantaneous reactive power ($Q$) in accordance with the depth of the voltage sag and the specific grid code in order to be able to improve the profile of the utility grid voltages.

3) To limit the amplitude of the three-phase currents through the inverter in order to avoid its disconnection from the utility grid due to the activation of the overcurrent protection.

Several strategies are described in the scientific literature in order to compute the PS and the NS components of unbalanced three-phase systems, as well as the system frequency and/or the phase angle. The Positive- negative- sequence detector (PNSD) [12,13] together with the synchronous reference frame (SRF) Phase-locked loop (PLL) [14] is used in $dq$ axes. In addition, a decoupled doubled SRF PLL [15], which is frequency- adaptive, can also be implemented to calculate the PS and the NS components when unbalanced voltage sags happen. Several synchronization algorithms are also described in the stationary reference frame ($\alpha\beta$ axes). Among them, the Dual second order generalized integrator frequency- locked loop (DSOGI-FLL) [16] and the Multiple second order generalized integrator frequency- locked loop (MSOGI-FLL) [17] can also calculate the PS and the NS components of the grid voltages, as well as its actual frequency. The latter is capable to remove the harmonic distortions from the calculated fundamental utility grid voltages.

When unlikely balanced voltage faults happen in the utility grid, only the PS component of the grid voltages is present, making the inverter currents have also its PS component. In this case, the instantaneous active power ($P$) and $Q$ will be constant values. However, the grid voltages at the point of common coupling (PCC) will have both the PS and the NS components for unbalanced faults, making not only $P$ and $Q$ oscillate with twice the system frequency ($2\omega_0$´) around its average value, but also the DC link voltage ($v_{DC}$). So, the link capacitor can be damaged if the amplitude of such oscillations is high, or can compromise the stability of the entirely grid-connected PV system unless the NS component of the inverter currents were delivered to the utility grid according to several strategies described in the scientific literature [13,18–21].

The LVRT capability is achieved for a two-stage grid-connected PV system in [7] in which an adaptive PLL is used to calculated the phase angle of the PS component of the three-phase grid voltages, meanwhile a Second order generalized integrator (SOGI) [16] detects the voltage sags and its depth. The main drawback of this strategy is the non-frequency- adaptive behavior of the two Proportional resonant (PR) current regulators.

Huka et. al [9] propose a LVRT algorithm for a two-stage grid-connected PV system in which a classical three-phase PLL is used and a Sequence separation method will extract the PS and the NS components of the utility grid voltages and currents so as to be able to control the currents with conventional Proportional integral (PI) regulators in $dq$ axes. In this case, a frequency-adaptive behavior is ensured by the feedback of the phase angle of the PS component to the PI- based current controllers, although the implementation of the control strategy must be carried out by using two different SRFs for both the PS and the NS components of the three-phase inverter currents [22], using four PI regulators and two notch filters.

Todorović et. al [10] and Cupertino et. al [11] use a DSOGI and also extract the PS and the NS components of the grid voltages and currents. The former controls a single-stage grid-connected generic system, whereas the latter controls a two-stage grid-connected PV system. In both cases, the current control is carried out with two frequency-adaptive PR current regulators in $\alpha\beta$ axes, although $\omega_0$´ might be distorted if the utility grid voltages had harmonics.

Díaz et al. estimate the system frequency in [21] by using a conventional PLL, as well as the PS and the NS components of the grid voltages using a Delayed signal cancellation (DSC) algorithm. However, the amplitudes of the three-phase inverter currents are not limited during the voltage sags and a poor implementation of the LVRT capability is remarkable.

Nowadays, the DSOGI-FLL synchronization algorithm mentioned above has been commonly used to calculate the PS and the NS components of the three-phase grid voltages, as well as $\omega_0$´. However, this algorithm cannot estimate those variables properly because the grid voltages at the PCC could be distorted with harmonics disserving the proper behavior and the stability of the system.

The drawbacks of the DSOGI-FLL can be overcome by implementing the MSOGI-FLL synchronization algorithm [17], which behaves as a frequency- adaptive scheme, calculates the PS and the NS components of the grid voltages at the fundamental frequency with no harmonic distortions, and has a better transient response [23] than the synchronization algorithm described in [15].

The main goal of this article is the development of the proper control of $P$ and $Q$ for a three-phase single-stage grid-connected PV system when balanced and unbalanced grid voltage sags happen together with the application of the LVRT ancillary service. This control is achieved by regulating the PS and the NS components of the three-phase inverter currents in the stationary reference frame ($\alpha\beta$ axes) with only two PR current controllers, allowing the use of less computational resources. In addition, a Harmonic Compensator (HC) structure can be cascaded to the PR current regulators in order to cancel the harmonic distortions of the grid currents produced by distorted utility grid voltages. The estimated undistorted system frequency at the output of the MSOGI-FLL synchronization algorithm is fed back to the resonant filters of both the PR current regulators and the HC structure, making the grid-connected PV system frequency- adaptive.

A new LVRT ancillary service is proposed according to the voltage profile recommended by the IEC 61400-21 normative [21,24], whereas the injected $Q$ to the utility grid will obey the specifications in accordance with the depth of the voltage sag in Spain [25]. The novelty of this proposal relies in the possibility of delivering a certain amount of constant $P$ to the utility grid when the depth of the voltage sag is small, as well as the strategy exerted to impose a limitation to the amplitude of the three-phase inverter currents to its nominal value in order to avoid the triggering of the overcurrent protection of the inverter and its disconnection from the mains during the voltage sags. This approach enhances the specific proposal made by Afshari et al. in [25] and Camacho et al. in [26].

The rest of the article is organized as follows: Section 2 describes the power and control subsystems of the three-phase single-stage grid-connected PV system, in which the MSOGI-FLL synchronization algorithm and the Current references Generator are presented. The control strategy is described in Section 3, where the PR current regulators, the HC structure, the limitation imposed to the amplitude of the three-phase currents for normal and faulty operation modes, and the development of the LVRT ancillary service during voltage sags are studied and implemented. A realized case study is presented in Section 4 where the grid-connected 500 kVA PV generator together with the parameters for the power and control subsystems are shown. Several simulations with MATLAB/SIMULINK and experiments using the Controller hardware-in-the loop (CHIL) Simulation technique are done in Sections 5 and 6, respectively. Finally, the conclusions of the paper are summarized in Section 7.

## 2. Overview of the single-stage grid-connected photovoltaic system

The block diagram of the power and control subsystems of the three-phase grid-connected PV system is shown in Fig. 1, and the lack of a boost DC/DC converter in this configuration is known as a single-stage scheme.

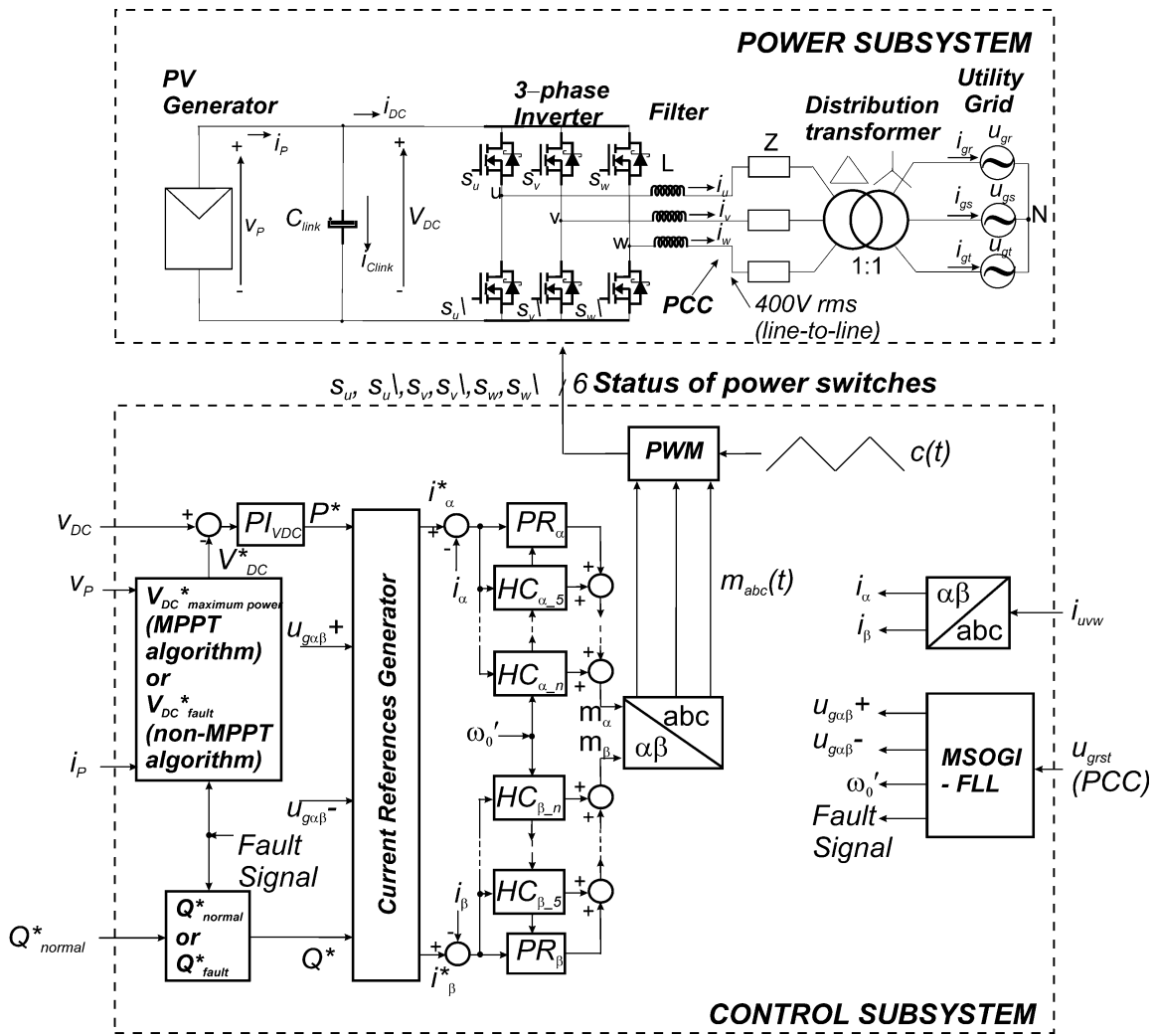

Fig. 1. Block diagram of the frequency- adaptive grid-connected PV system in the stationary reference frame control.

The power subsystem is built with the PV generator, a three-phase Voltage source inverter (VSI) as the power conditioner, the link capacitor ($C_{link}$), the line inductance of the filter ($L$), the grid impedance ($Z$), the distribution transformer, and the low-voltage utility grid. The grid voltages ($u_{grst}$), the inverter currents ($i_{uvw}$) in the AC side, as well as the DC link voltage ($v_{DC}$), and the current at the output of the PV generator ($i_P$) in the DC side are measured and fed back to the control subsystem in order to do the power balance between the PV generator and the utility grid. The voltage $v_{DC}$ is the same as the DC voltage at the output of the PV generator ($v_P$). The DC current at the input of the inverter is $i_{DC}$ which approximately equals $i_P$ at steady-state for a large link capacitor in which the DC current ($i_{Clink}$) can be neglected. It is worth noting that the grid currents ($i_{grst}$) equal the inverter currents ($i_{uvw}$) because the transformation ratio of the distribution transformer is 1:1. Hence, both variables will be used indistinctly throughout the paper referring to the controlled currents. In addition, the utility grid voltages at the PCC are affected by other power sources because of the effect produced by $Z$, which is the equivalent Thèvenin impedance seen by the grid-connected PV system, and is calculated by dividing the open circuit grid voltage by the short circuit current at the PCC ($Z = V_{oc}/I_{sc}$). If $Z$ is mainly resistive with a large amplitude, the utility grid will behave as a weak grid; if it has a small amplitude and is mainly inductive, a stiff (strong) grid is attained. The amplitude and the frequency of the voltage at PCC for stiff grids are less affected by load variations than for weak grids. Hence, the lattter worsens the profile of the grid voltages.

The control subsystem includes the synchronization algorithm, the Current References Generator (CRG), the inner current controller together with the outermost voltage controller for a cascaded control scheme, the Maximum Power Point Tracking (MPPT) and non-MPPT algorithms [25,27–30] in order to implement the LVRT ancillary service when voltage sags happen, and the Pulse Width Modulation (PWM) blocks. The Perturb and Observe (P&O) method has been used in this paper to track the MPP during normal operation and the non-MPP during voltage sags.

## 2.1. Synchronization algorithm

The MSOGI-FLL [17] synchronization algorithm together with the Fault detector algorithm, whose block diagram is depicted in Fig. 2, are proposed in this paper.

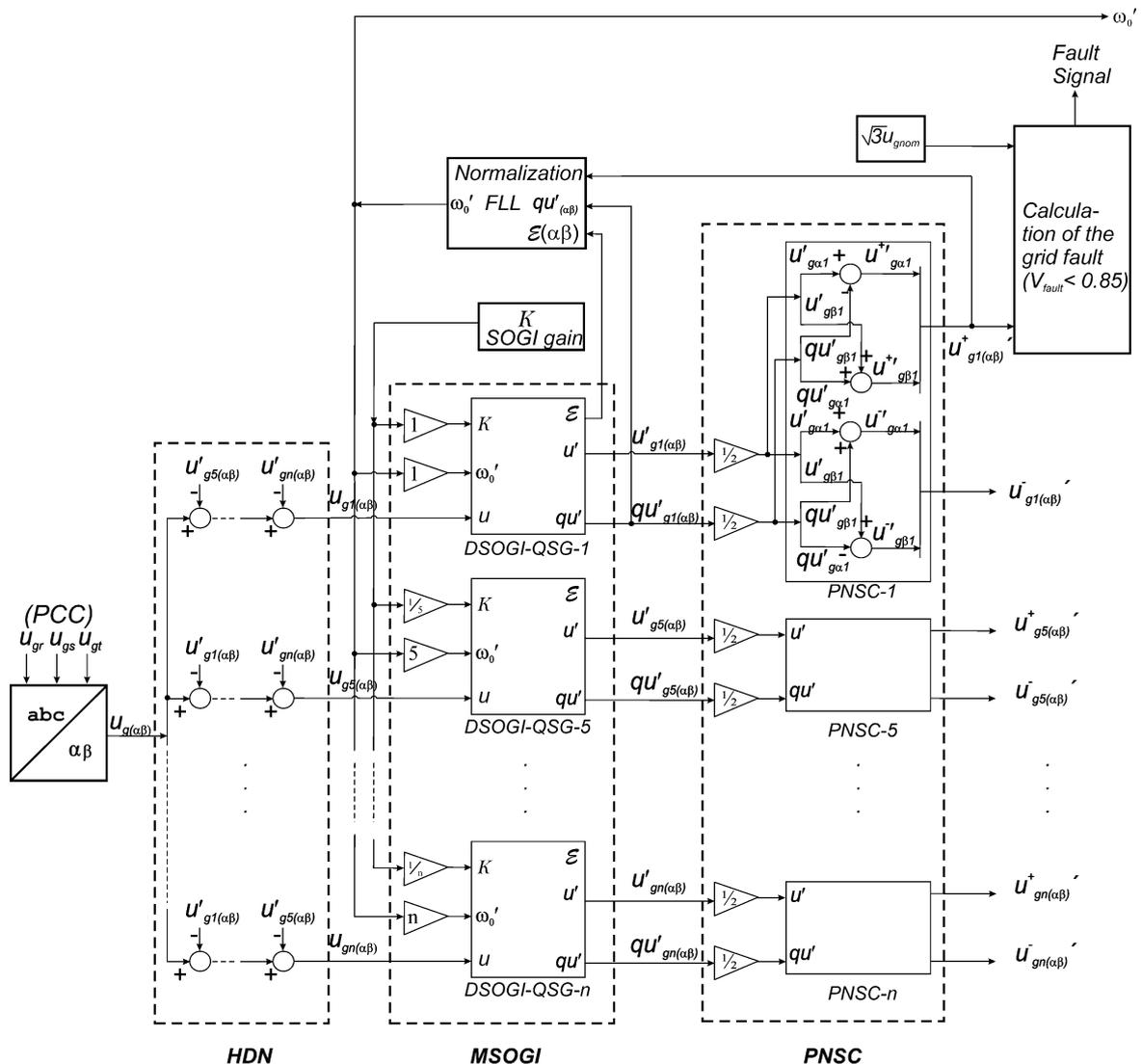

Fig. 2. Block diagram of the MSOGI-FLL synchronization algorithm [17] + Fault Detection.

The use of this synchronization algorithm estimates the system frequency $\omega_0'$ of the grid voltages without distortions due to the effect of the Harmonic decoupling network (HDN) and outputs their PS and NS components. The HDN is fed by the $\alpha\beta$ quantities of the instantaneous three-phase grid voltages (phase-to-neutral) at PCC, which are calculated applying the Clarke´s transformation [31]:

$$\begin{bmatrix} u_{g\alpha} \\ u_{g\beta} \end{bmatrix} = \sqrt{\frac{2}{3}} \begin{bmatrix} 1 & -\frac{1}{2} & -\frac{1}{2} \\ 0 & \frac{\sqrt{3}}{2} & -\frac{\sqrt{3}}{2} \end{bmatrix} \begin{bmatrix} u_{gr} \\ u_{gs} \\ u_{gt} \end{bmatrix} \tag{1}$$

and selects the proper harmonics of the three-phase grid voltages. These harmonics are delivered to two Dual second order generalized integrators with quadrature signal generator (DSOGI-QSG-$i$) where $i=1,5,…$n are the subscripts corresponding to the fundamental, the $5^{th}$ and the $n^{th}$ harmonics respectively, and $q = e^{-j\frac{\pi}{2}}$ is a -90° shift parameter to obtain the quadrature signal of the waveform at the input. The estimation of the system frequency ($\omega_0'$) is made by the FLL block described in Fig. 3 (b), multiplied by the $i^{th}$ harmonic, and fed back to the corresponding DSOGI-QSG-$i$ block described in Fig. 3 (a), whose outputs are delivered to the Positive-negative- sequence calculator for each harmonic (PNSC-$i$).

The DSOGI-QSG-$i$ block behaves as an adaptive resonant filter in a generalized integrator (GI) for sinusoidal signals scheme [32], where the resonance frequency is precisely the frequency of the harmonic signal at its input ($i \times \omega_0'$). The transfer functions $\frac{u'}{u}(s)$ and $\frac{qu'}{u}(s)$ ($qu'$ in quadrature with the input signal $u'$) correspond to a band-pass and a low-pass filters respectively, and its main behavior only depends of the value of gain $K$, which is calculated according to the proper tradeoff between the dynamic response and the harmonic rejection, attaining $K = \sqrt{2}$. Regarding the tuning of the FLL algorithm, the normalization of the FLL gain linearizes the system which will not depend neither of the grid voltages nor of the gain $K$ of the DSOGI-QSG block, but only of gain $\Gamma$ [17].

It is worth noting that $u^+_{gI(\alpha\beta)}'$ and $u^-_{gI(\alpha\beta)}'$ are the estimation of the PS and the NS $\alpha\beta$ quantities of the grid voltages at $\omega_0'$.

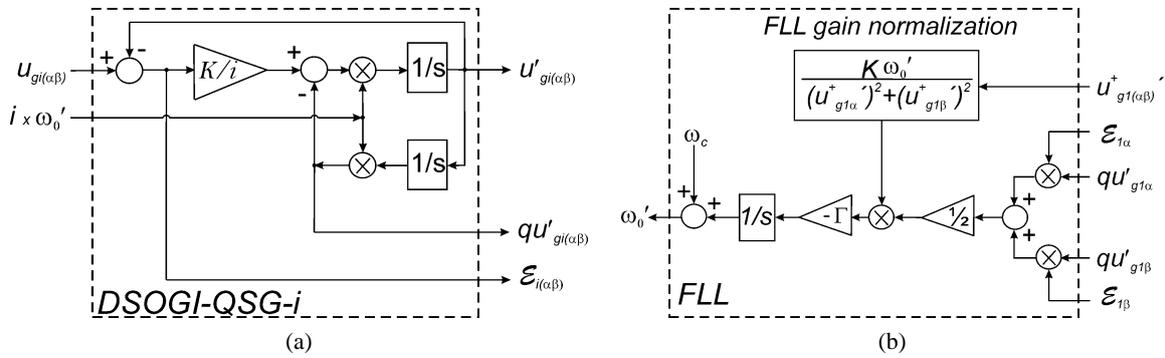

Fig. 3. Block diagram of the DSOGI-QSG-i and the FLL: (a) DSOGI-QSG-i, (b) FLL.

Finally, the *Fault Signal* flag is activated when the normalized depth of the voltage sag ($V_{fault}$) calculated by eq. (7) is less than 0.85 according to IEC 61400-21.

## 2.2. Current References Generator

The main goal of the CRG is to calculate the reference current commands according to the desired behavior, i.e., constant and/or oscillations at $2\omega_0'$ of $P$ and $Q$, balanced or unbalanced sinusoidal currents, etc. The stationary reference frame will be used throughout this paper and $P$, $Q$ can be described as follows in $\alpha\beta$ axes [21,25,26]:

$$P = P_0^+ + P_0^- + \tilde{P}$$
$$P_0^+ = u_{g\alpha}^+ i_\alpha^+ + u_{g\beta}^+ i_\beta^+ \tag{2}$$
$$P_0^- = u_{g\alpha}^- i_\alpha^- + u_{g\beta}^- i_\beta^-$$

$$\tilde{P} = u_{g\alpha}^+ i_\alpha^- + u_{g\beta}^+ i_\beta^- + u_{g\alpha}^- i_\alpha^+ + u_{g\beta}^- i_\beta^+$$

$$Q = Q_0^+ + Q_0^- + \tilde{Q}$$

$$Q_0^+ = u_{g\beta}^+ i_\alpha^+ - u_{g\alpha}^+ i_\beta^+$$

$$Q_0^- = u_{g\beta}^- i_\alpha^- - u_{g\alpha}^- i_\beta^-$$

$$\tilde{Q} = u_{g\beta}^+ i_\alpha^- - u_{g\alpha}^+ i_\beta^- + u_{g\beta}^- i_\alpha^+ - u_{g\alpha}^- i_\beta^+$$

$$(3)$$

where $P_0^+ + P_0^-$ ($Q_0^+ + Q_0^-$) are the average active (reactive) powers with the PS and the NS components respectively, $\tilde{P}$ ($\tilde{Q}$) is the oscillating active (reactive) power at $2\omega_0{'}$, and $P$ ($Q$) is the instantaneous active (reactive) power.

There are infinite combinations in order to obtain the proper values of $P_0$, $Q_0$, $\tilde{P}$ and $\tilde{Q}$, as well as the nature of the three-phase currents delivered to the utility grid. For example, Rodríguez et. al [13] present five possibilities to deal with the behavior of $P$ and $Q$, as well as the sinusoidal nature of the three-phase currents when one-phase to ground faults happen (unbalanced fault):

1) $P$ and $Q$ are DC constants, producing distorted non-sinusoidal three-phase currents.
2) $P$ is constant and $Q$ has light oscillations at $2\omega_0{'}$, producing light distorted non-sinusoidal three-phase currents.
3) $P$ is constant and $Q$ has strong oscillations at $2\omega_0{'}$, producing unbalanced sinusoidal three-phase currents.
4) $Q$ is constant and $P$ has strong oscillations at $2\omega_0{'}$, producing unbalanced sinusoidal three-phase currents.
5) Both, $P$ and $Q$ have oscillations at $2\omega_0{'}$, producing balanced sinusoidal three-phase currents.

In order to exert a constant active power control to protect the link capacitor from large oscillations at $2\omega_0{'}$ which can damage it due to thermal stress, and inject sinusoidal three-phase currents, the third condition mentioned above imposes that $\tilde{P}$ must be set to zero in eq. (2).

The CRG block is fed by the active ($P*$), the reactive ($Q*$) power reference commands, and the PS and NS of the $\alpha\beta$ quantities of the three-phase utility grid voltages ($u_{g\alpha\beta}^+$ and $u_{g\alpha\beta}^-$) that are calculated by the MSOGI-FLL block, and delivers the current reference commands in $\alpha\beta$ axes of the three-phase inverter currents to the two PR regulators.

As said before, both the PS and the NS components of the inverter current reference commands in $\alpha\beta$ axes ($i_{\alpha\beta}^{+*}$ and $i_{\alpha\beta}^{-*}$) play a key role in the behavior of the grid-connected inverter. So, the outputs of the CRG can be expressed as:

$$i_\alpha^* = i_\alpha^{+*} + i_\alpha^{-*} = i_{\alpha P}^* + i_{\alpha Q}^*$$

$$i_\beta^* = i_\beta^{+*} + i_\beta^{-*} = i_{\beta P}^* + i_{\beta Q}^*$$

$$(4)$$

where $i_{\alpha\beta P}^*, i_{\alpha\beta Q}^*$ are the current references according to $P*$ and $Q*$ respectively.

Afshari et. al [25] demonstrates that the third possibility mentioned above for one-phase to ground faults "exploits the full capacity" of the inverter for constant active power operation when unbalanced faults happen. For this, the right part of eq. (4) can be expressed as follows:

$$i_{\alpha P}^* = \frac{u_{g\alpha}^+ - u_{g\alpha}^-}{\left( u_{g\alpha}^{+^2} + u_{g\beta}^{+^2} \right) - \left( u_{g\alpha}^{-2} + u_{g\beta}^{-2} \right)} P*$$

$$i_{\alpha Q}^* = \frac{u_{g\beta}^+ + u_{g\beta}^-}{\left( u_{g\alpha}^{+^2} + u_{g\beta}^{+^2} \right) - \left( u_{g\alpha}^{-2} + u_{g\beta}^{-2} \right)} Q*$$

$$(5)$$

$$i_{\beta P}^* = \frac{u_{g\beta}^+ - u_{g\beta}^-}{\left(u_{g\alpha}^{+2} + u_{g\beta}^{+2}\right) - \left(u_{g\alpha}^{-2} + u_{g\beta}^{-2}\right)} P^*$$

$$i_{\beta Q}^* = -\frac{u_{g\alpha}^+ + u_{g\alpha}^-}{\left(u_{g\alpha}^{+2} + u_{g\beta}^{+2}\right) - \left(u_{g\alpha}^{-2} + u_{g\beta}^{-2}\right)} Q^*$$

It is worth noting in eq. (5) that if balanced faults happen for an *abc* (*acb*) sequence- system, the NS (PS) component of the grid voltages and currents in $\alpha\beta$ axes equals zero, attaining only the PS (NS) component of the currents.

## 3. Implementation of the control strategy

The outputs of the CRG block are the PS and the NS components of the grid current reference commands ($i^*_{\alpha\beta}$) which are compared with the real currents ($i_{\alpha\beta}$). The error is fed to two Proportional Resonant regulators (PR$_\alpha$ and PR$_\beta$) and two Harmonic compensators (HC$_{\alpha\_5,...n}$ and HC$_{\beta\_5,...n}$) described by eq. (6) which are used in this case to control the $\alpha\beta$ quantities of the three-phase inverter currents in the innermost loop [33]. The former regulates the $\alpha\beta$ quantities of the three-phase inverter currents, whereas the latter compensate the presence of the harmonics of the grid voltages as perturbations in the current loops. A PI regulator (PI$_{VDC}$) is used to control the DC link voltage $v_{DC}$ in the outer loop [34] whose output is $P^*$, in a cascaded control structure. The DC link voltage reference command ($V^*_{DC}$) is the output of the MPPT and the non-MPPT algorithms module for normal and faulty operation modes, respectively.

It is worth noting that in addition to $P^*$ and $Q^*$, the CRG block is fed by the incoming $\alpha\beta$ quantities of the instantaneous PS and NS components of the three-phase grid voltages. Moreover, the MPPT and non-MPPT blocks together with the reactive power reference block are fed by the *Fault Signal* flag which is delivered by the Synchronization algorithm block in order to detect the voltage sags.

$$PR_{\alpha\beta}(s) = K_{P\alpha\beta} + \frac{2K_{I\alpha\beta}\omega_c s}{s^2 + 2\omega_c s + \omega_o'^2}$$

$$HC_{\alpha\beta\_5}(s) = \frac{2K_{I\alpha\beta\_5}\omega_c s}{s^2 + 2\omega_c s + (5\omega_0')^2}$$

$$.$$
$$.$$
$$.$$

$$HC_{\alpha\beta\_n}(s) = \frac{2K_{I\alpha\beta\_n}\omega_c s}{s^2 + 2\omega_c s + (n\omega_0')^2}$$

(6)

$K_{P\alpha\beta}$ and $K_{I\alpha\beta}$ are the gains of the transfer function PR$_{\alpha\beta}$(s) of the PR regulators, $\omega_o'$ is the resonance angular frequency which equals the estimated system frequency of the three-phase utility grid voltages, and $\omega_c$ is the cut-off frequency. $K_{I\alpha\beta\_5,...n}$ are the integral gains of the transfer function HC$_{\alpha\beta\_5,...n}$(s) of the HC structure.

The outputs of the PR regulators and the HC structure are the reference modulating signals $m_\alpha$ and $m_\beta$, which are delivered to the PWM module after the application of the inverse Clarke´s transformation [31] producing the modulating signal $m_{abc}(t)$ that is compared with the triangular carrier signal $c(t)$ in the PWM module in order to trigger the SiC MOSFET-based switches (ON-OFF operation) of the three-phase inverter.

It is worth noting that the feedback of $\omega_0'$ estimated by the MSOGI-FLL block to the PR regulators and $i \times \omega_0'$ to the corresponding resonant filter of the HC structure (i=5, 5, 7, 11, 13,…n) makes the system frequency-adaptive, allowing the inverter currents to be free of harmonic distortions as well.

The output of PI$_{VDC}$ delivers $P^*$ which is proportional to the incoming power from the PV generator. The value of $Q^*$ is configured as an open loop scheme in accordance with the depth of the voltage sag and will be able to control the *PF* of the system, which will be explained in the next topic.

### 3.1. Control of the three-phase VSI during voltage sags in the stationary reference frame

The international grid codes [19,25] are very restrictive when severe grid faults happen. During very short duration faults, each country has its own normative, although most of them impose the inverter not only to remain connected to the utility grid, but also to deliver some $Q$ in accordance with the depth of the voltage sag so as to be able to improve the voltage profile of the grid voltages. Moreover, a limitation in the amplitude of the output three-phase line currents through the inverter to its nominal values is mandatory in order to avoid the triggering of the overcurrent protection and its disconnection from the grid.

Both, the injection of the adequate $Q$ during voltage sags and the limitation in the amplitude of the output three-phase line currents through the inverter are known as LVRT capability.

It is remarkable that when the depth of the voltage sag is not severe, some amount of $P$ can also be delivered to the utility grid.

Figures 4 (a) and (b) depict the LVRT requirements according to IEC 61400-21 [21,24] and the allowed $Q$ in accordance with the normalized depth of the voltage sag in Spain [25], respectively.

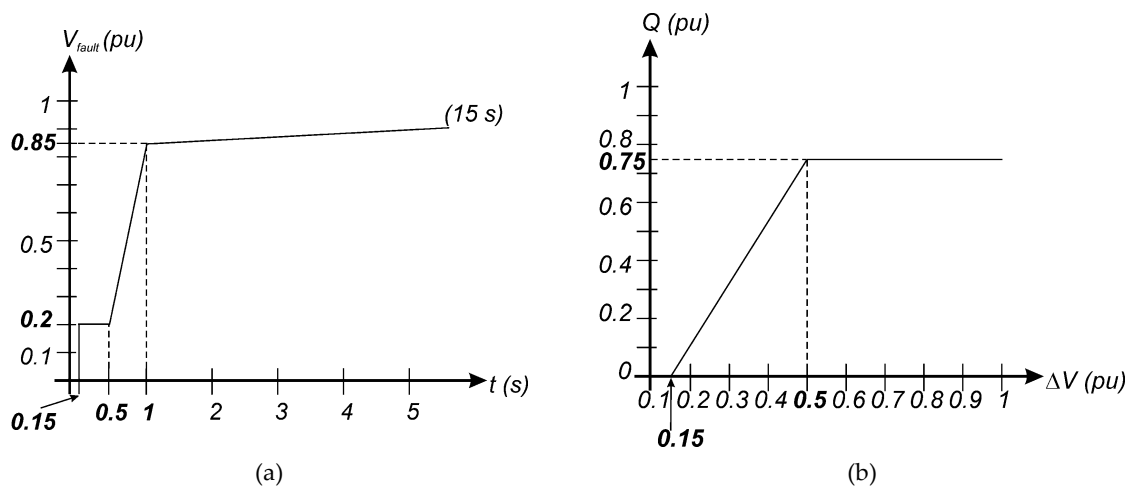

(a)             (b)

Fig. 4. (a) Voltage profile for the LVRT capability (IEC 61400-21) [21,24], (b) $Q$ delivered to the utility grid during voltage sags in Spain ($\Delta V(pu) = 1 - Vfault$) [25].

In order to implement the LVRT capability, the control algorithm must detect the depth of the voltage sag in a time duration much shorter than the allowed of this according to the recommended grid code. So, the normalized depth of the voltage sag ($V_{fault}$) is calculated by measuring the three-phase grid voltages, extracting its PS component in the stationary reference frame, calculating its amplitude, and normalizing with the *root-mean-square (rms)* value of the nominal line-to-line utility grid voltages [21,25,35] as shown in eq. (7).

$$V_{fault} = \frac{\sqrt{u_{g\alpha}^{+2} + u_{g\beta}^{+2}}}{\sqrt{3}u_{gnom}} \qquad (7)$$

where $u_{gnom}$ is the nominal *rms value of* the phase-to neutral grid voltages. On one hand, eq. (7) shows that when normal operation mode is activated, the numerator reaches its maximum value and $V_{fault} = 1$. On the other hand, for faulty operation mode, the amplitude of the PS component of the grid voltages decreases and $V_{fault} < 1$. The IEC 61400-21 imposes that a voltage sag has occurred if $V_{fault} < 0.85$.

The maximum apparent power during the faulty operation mode ($S_{fault}$) will be calculated with the amplitudes of the PS and the NS components of the grid voltages, using the value of the nominal apparent power $S_{nom}$, and normalizing again with the *rms* value of the nominal line-to-line utility grid voltages:

$$S_{fault} = \frac{\sqrt{u_{g\alpha}^{+2} + u_{g\beta}^{+2}} - \sqrt{u_{g\alpha}^{-2} + u_{g\beta}^{-2}}}{\sqrt{3}u_{gnom}} S_{nom} \qquad (8)$$

where the maximum nominal apparent power $S_{nom}$ is about 500 kVA (see Table 1). For example, when no voltage sag is detected, the right part of the numerator is zero and $S_{fault} = S_{nom}$; on the contrary when balanced faults happen, no NS component of the grid voltages is present and the PS component decreases; finally, for unbalanced faults, the NS component increases and the PS component decreases. The last two situations imply that $S_{fault} < S_{nom}$ and a limitation in the amplitude of the three-phase inverter currents is achieved, as will be explained in detail in Topic 3.2.

$Q$ is calculated in accordance with the depth of the voltage sag in eq. (7) and Fig. 4 (b), as follows:

$$\left\{ \begin{array}{ll} Q = 0, & V_{fault} \geq 0.85 \\ Q = \frac{15}{7} S_{nom}\big(0.85 - V_{fault}\big), & 0.5 \leq V_{fault} < 0.85 \\ Q = \frac{3}{4} S_{nom}, & V_{fault} < 0.5 \end{array} \right. \tag{9}$$

According to $S_{fault}$ and $Q$, a certain amount of instantaneous active power $P_{fault}$ can be delivered to the grid in the faulty operation mode, keeping the limitation in the amplitude of the currents [36]:

$$P_{fault} = \sqrt{S_{fault}^{\,2} - Q^2} \tag{10}$$

### 3.2. Limitation in the amplitude of the three-phase inverter currents

Defining the in-quadrature components of the $\alpha\beta$ quantities of the three-phase grid voltages $u_{g\alpha} \perp = -u_{g\beta}$ and $u_{g\beta} \perp = u_{g\alpha}$ [25], eq. (5) can be re-written as follows:

$$i_P^* = i_{\alpha P}^* + i_{\beta P}^* = \frac{K_P\big(u_{g\alpha}^+ + u_{g\beta}^+\big)}{\big(u_{g\alpha}^{+\,2} + u_{g\beta}^{+\,2}\big) - 2K_P\big(u_{g\alpha}^{-\,2} + u_{g\beta}^{-\,2}\big)} P^*$$

$$- \frac{(1 - K_P)\big(u_{g\alpha}^- + u_{g\beta}^-\big)}{\big(u_{g\alpha}^{-\,2} + u_{g\beta}^{-\,2}\big) - 2(1 - K_P)\big(u_{g\alpha}^{-\,2} + u_{g\beta}^{-\,2}\big)} P^*$$

$$i_Q^* = i_{\alpha Q}^* + i_{\beta Q}^* \tag{11}$$

$$= -\frac{K_Q(u_{g\alpha\perp}^+ + u_{g\beta\perp}^+)}{\big(u_{g\alpha}^{+\,2} + u_{g\beta}^{+\,2}\big) - 2K_Q\big(u_{g\alpha}^{-\,2} + u_{g\beta}^{-\,2}\big)} Q^*$$

$$- \frac{(1 - K_Q)(u_{g\alpha\perp}^- + u_{g\beta\perp}^-)}{\big(u_{g\alpha}^{-\,2} + u_{g\beta}^{-\,2}\big) - 2(1 - K_Q)\big(u_{g\alpha}^{-\,2} + u_{g\beta}^{-\,2}\big)} Q^*$$

where $\frac{1}{K_P} = \frac{1}{K_Q} = 1 + \frac{u_{g\alpha}^{-2} + u_{g\beta}^{-2}}{u_{g\alpha}^{+2} + u_{g\beta}^{+2}}$ , $\frac{1}{1-K_P} = \frac{1}{1-K_Q} = 1 + \frac{u_{g\alpha}^{+2} + u_{g\beta}^{+2}}{u_{g\alpha}^{-2} + u_{g\beta}^{-2}}$ , $Q^* = Q$ according to eq. (9), $P^* = P_{fault}$ according to eq. (10) and $K_P = K_Q \in [0,1]$. $K_P = K_Q = 1$ for no faulty or balanced faulty operation modes with the positive- sequence (*abc*) of the three-phase utility grid voltages, $0 < K_P = K_Q < 1$ for unbalanced faulty operation mode, and $K_P = K_Q = 0$ for no faulty operation mode with the negative- sequence (*acb*) of the three-phase utility grid voltages.

The numerators of eq. (11) can be expressed in vector form as:
$u_{g\alpha}^+ + ju_{g\beta}^+ = \mathbf{u_g^+}$, $u_{g\alpha\perp}^+ + ju_{g\beta\perp}^+ = \mathbf{u_{g\perp}^+}$, $u_{g\alpha}^- + ju_{g\beta}^- = \mathbf{u_g^-}$, $u_{g\alpha\perp}^- + ju_{g\beta\perp}^- = \mathbf{u_{g\perp}^-}$,

and the denominators are simplified as:

$\left(u_{g\alpha}^{+}{}^2 + u_{g\beta}^{+}{}^2\right) - 2K_P\left(u_{g\alpha}^{-}{}^2 + u_{g\beta}^{-}{}^2\right) = |\mathbf{u'_g^+}|^2$, $\left(u_{g\alpha}^{-}{}^2 + u_{g\beta}^{-}{}^2\right) - 2(1 - K_P)\left(u_{g\alpha}^{-}{}^2 + u_{g\beta}^{-}{}^2\right) = |\mathbf{u'_g^-}|^2$, knowing that $K_P = K_Q$.

Then eq. (11) can be expressed as:

$$\mathbf{i_P^*} = \frac{K_P P^*}{|\mathbf{u'_g^+}|^2}\mathbf{u_g^+} - \frac{(1 - K_P)P^*}{|\mathbf{u'_g^-}|^2}\mathbf{u_g^-}$$

$$\mathbf{i_Q^*} = -\frac{K_Q Q^*}{|\mathbf{u'_g^+}|^2}\mathbf{u_{g\perp}^+} - \frac{(1 - K_Q)Q^*}{|\mathbf{u'_g^-}|^2}\mathbf{u_{g\perp}^-}$$

(12)

where:

$\mathbf{i_P^*}$ and $\mathbf{i_Q^*}$ are the reference inverter current vectors according to $P^*$ and $Q^*$.

$\mathbf{u_g^+}$ and $\mathbf{u_{g\perp}^+}$ are the positive- sequence component of the grid voltage vector and its in-quadrature component.

$\mathbf{u_g^-}$ and $\mathbf{u_{g\perp}^-}$ are the negative- sequence component of the grid voltage vector and its in-quadrature component.

$|\mathbf{u'_g^+}|^2$ and $|\mathbf{u'_g^-}|^2$ are the equivalent square of the amplitude of the PS and the NS components of the grid voltage vectors.

As can be seen in eq. (12), for unbalanced voltage sags, adding (or subtracting) the PS and the NS components of the grid voltage vectors results in an *ellipse* in $\alpha$ and $\beta$ axes [23] for $\mathbf{i_P^*}$ and $\mathbf{i_Q^*}$, respectively. So, the phasors of the instantaneous inverter currents in the complex plane can be expressed as follows:

$$\mathbf{i_P^*} = \frac{K_P P^*}{|\mathbf{u'_g^+}|^2}|\mathbf{u_g^+}|e^{j\omega_0 t} - \frac{(1 - K_P)P^*}{|\mathbf{u'_g^-}|^2}|\mathbf{u_g^-}|e^{-j\omega_0 t}$$

$$\mathbf{i_Q^*} = -\left[\frac{K_Q Q^*}{|\mathbf{u'_g^+}|^2}|\mathbf{u_g^+}|e^{j\omega_0 t} + \frac{(1 - K_Q)Q^*}{|\mathbf{u'_g^-}|^2}|\mathbf{u_g^-}|e^{-j\omega_0 t}\right]e^{j\frac{\pi}{2}}$$

(13)

Operating with eq. (13):

$$\mathbf{i_P^*} = \left[\frac{K_P P^*}{|\mathbf{u'_g^+}|^2}|\mathbf{u_g^+}| - \frac{(1 - K_P)P^*}{|\mathbf{u'_g^-}|^2}|\mathbf{u_g^-}|\right]cos\omega_0 t$$

$$+ j\left[\frac{K_P P^*}{|\mathbf{u'_g^+}|^2}|\mathbf{u_g^+}| + \frac{(1 - K_P)P^*}{|\mathbf{u'_g^-}|^2}|\mathbf{u_g^-}|\right]sin\omega_0 t$$

(14)

$$\mathbf{i_Q^*} = \left[\frac{K_Q Q^*}{|\mathbf{u'_g^+}|^2}|\mathbf{u_g^+}| - \frac{(1 - K_Q)Q^*}{|\mathbf{u'_g^-}|^2}|\mathbf{u_g^-}|\right]sin\omega_0 t$$

$$- j\left[\frac{K_Q Q^*}{|\mathbf{u'_g^+}|^2}|\mathbf{u_g^+}| + \frac{(1 - K_Q)Q^*}{|\mathbf{u'_g^-}|^2}|\mathbf{u_g^-}|\right]cos\omega_0 t$$

where:

$$I_{PLarge} = \frac{K_P P^*}{|\mathbf{u'_g^+}|^2}|\mathbf{u_g^+}| + \frac{(1-K_P)P^*}{|\mathbf{u'_g^-}|^2}|\mathbf{u_g^-}|, \quad I_{PShort} = \frac{K_P P^*}{|\mathbf{u'_g^+}|^2}|\mathbf{u_g^+}| - \frac{(1-K_P)P^*}{|\mathbf{u'_g^-}|^2}|\mathbf{u_g^-}|$$

and

$$I_{QLarge} = \frac{K_Q Q^*}{|\mathbf{u'_g^+}|^2}|\mathbf{u_g^+}| + \frac{(1-K_Q)Q^*}{|\mathbf{u'_g^-}|^2}|\mathbf{u_g^-}|, \quad I_{QShort} = \frac{K_Q Q^*}{|\mathbf{u'_g^+}|^2}|\mathbf{u_g^+}| - \frac{(1-K_Q)Q^*}{|\mathbf{u'_g^-}|^2}|\mathbf{u_g^-}|$$

Then, according to eq. (14):

$$\mathbf{i^*} = \mathbf{i_P^*} + \mathbf{i_Q^*} = \begin{bmatrix} i_\alpha^* \\ i_\beta^* \end{bmatrix} = \begin{bmatrix} i_{\alpha P}^* + i_{\alpha Q}^* \\ i_{\beta P}^* + i_{\beta Q}^* \end{bmatrix} = \begin{bmatrix} I_{PShort} cos\omega_0 t + I_{QShort} sin\omega_0 t \\ I_{PLarge} sin\omega_0 t - I_{QLarge} cos\omega_0 t \end{bmatrix} \tag{15}$$

where $\mathbf{i^*}$ is the reference inverter current vector.

Hence:

$$i_\alpha^* = K_\alpha \cos(\omega_0 t + \theta_\alpha); \; K_\alpha = \sqrt{I_{PShort}{}^2 + I_{QShort}{}^2}; \; \theta_\alpha = tan^{-1}(\frac{I_{QShort}}{I_{PShort}})$$

$$i_\beta^* = K_\beta \sin(\omega_0 t + \theta_\beta); \; K_\beta = \sqrt{I_{PLarge}{}^2 + I_{QLarge}{}^2}; \; \theta_\beta = tan^{-1}(\frac{I_{QLarge}}{I_{PLarge}})$$
$$\tag{16}$$

where $K_\alpha$ and $K_\beta$ are the amplitudes of the reference currents in $\alpha\beta$ axes, whereas $\theta_\alpha$ and $\theta_\beta$ are the initial phases for both reference currents.

In short, $I_{PLarge}$, $I_{QLarge}$, $I_{PShort}$, $I_{QShort}$ are the magnitudes of the large and the short axes of the corresponding ellipses in $\alpha\beta$ axes as the general case. There are four well-defined situations:

1) When no voltage sags happen, no reactive power is delivered to the utility grid ($Q = 0$) for unitary power factor operation, and the maximum available instantaneous active power $P$ is fed to the utility grid. So, $I_{PLarge} = I_{PShort} = I_{Pmax}$. In this case, $I_{QLarge} = I_{QShort} = 0$ and $K_\alpha = K_\beta = K_{max}$, yielding the maximum nominal value for the amplitude of the reference currents and describing a *circumference* in $\alpha\beta$ axes.

2) For three-phase deep and moderate voltage sags (balanced fault), $Q > S_{fault}$. According to eq. (10), non-real values are obtained and so, $P^* = P_{fault} = 0$ and $Q^* = S_{fault}$. In this case, $I_{QLarge} = I_{QShort} = I_{Qmax}$, whereas $I_{PLarge} = I_{PShort} = 0$ and also $K_\alpha = K_\beta = K_{max}$, yielding again the maximum nominal value for the amplitude of the reference currents and describing a *circumference* in $\alpha\beta$ axes, although $PF = 0$.

3) For three-phase small voltage sags (balanced fault), $Q < S_{fault}$ and according to eq. (10), $P^* = P_{fault} > 0$ and $Q^*$ has a small value. In this case, $I_{PLarge} = I_{PShort} < I_{Pmax}$, whereas $I_{QLarge} = I_{QShort}$ with a small value. However, $K_\alpha = K_\beta = K_{max}$, yielding again the maximum nominal value for the amplitude of the reference currents and describing a *circumference* in $\alpha\beta$ axes, although $PF > 0$.

4) For unbalanced voltage sags, the values of $I_{PLarge}$, $I_{PShort}$, $I_{QLarge}$ and $I_{QShort}$ are, in general, different from each other, and what is more important, $K_\alpha \neq K_\beta \leq K_{max}$, describing in this case an *ellipse* with less amplitudes than the maximum nominal value seen in the previous situations.

The application of the inverse Clarke´s transformation to the reference currents in $\alpha\beta$ axes for the four situations seen before guarantees the limitation of the three-phase inverter currents to its nominal values, with a maximum in the cases a *circumference* were described in $\alpha\beta$ axes.

Next, the previous four situations are further explained through the implementation of the LVRT capability.

### 3.3. Implementation of the low-voltage ride-through

The LVRT ancillary service of the control algorithm is described with the flowchart depicted in Fig. 5, where the normal or faulty operation modes are enabled according to the normalized depth of the voltage sag ($V_{fault}$) [37].

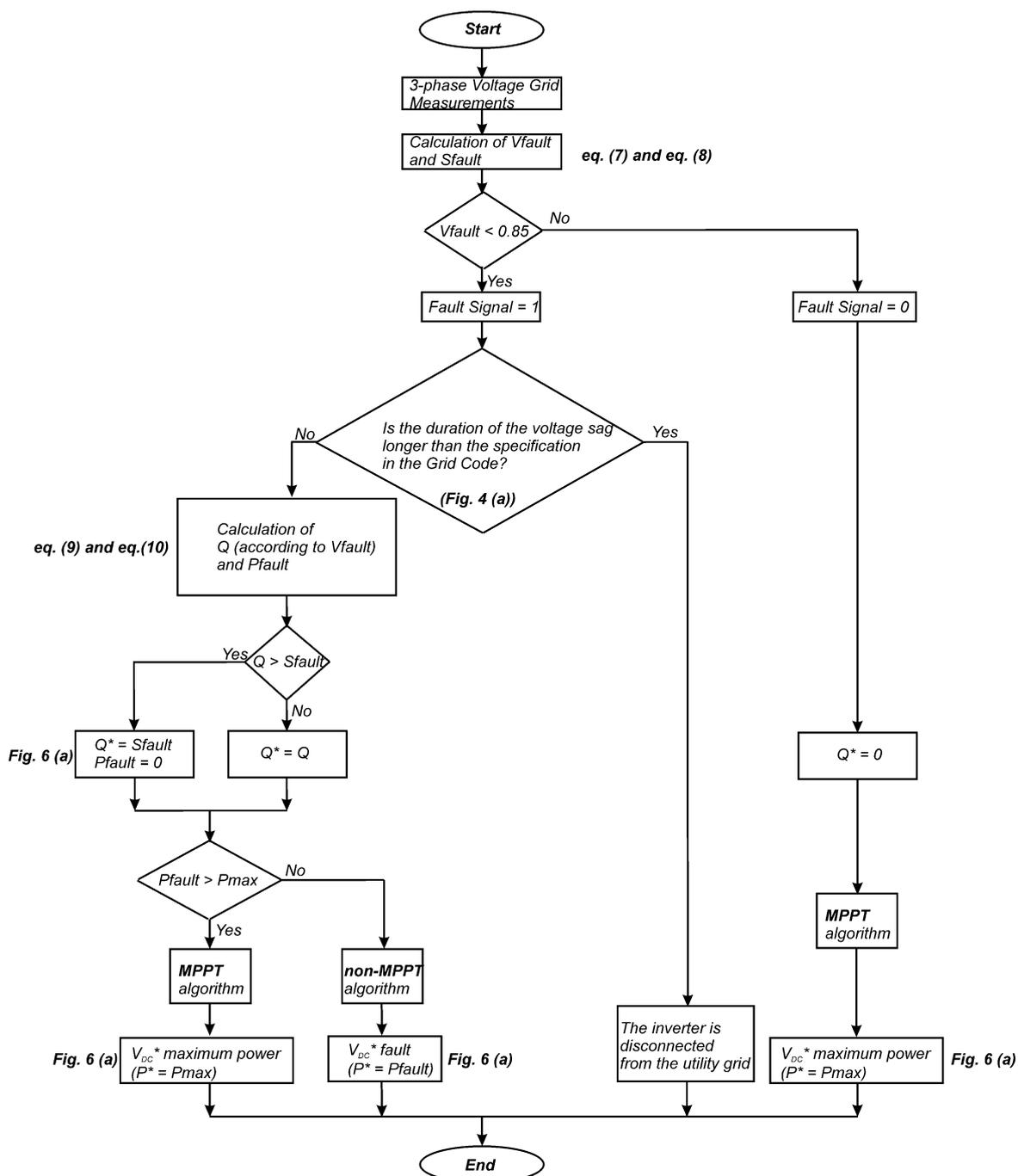

Fig. 5. Flowchart of the LVRT ancillary service of the control algorithm.

The normal operation mode is enabled when $V_{fault} \geq 0.85$ as said before. In this case, the *Fault Signal* flag is set to zero indicating that no fault has been detected and an unity power operation is established ($Q^* = 0$). It will activate the MPPT algorithm to inject all the available power at the output of the PV generator ($P^* = P_{max}$) for a specific irradiance and temperature.

On the contrary, the faulty operation mode is enabled when $V_{fault}$ is less than 0.85, activating the *Fault Signal* flag (*Fault Signal = 1*) and the non-MPPT algorithm in order to deliver less active power ($P^* = P_{fault}$). In this case, there are two possibilities: (i) the disconnection of the inverter from the utility grid or (ii) the activation of the LVRT capability. The former deals with a long duration of the voltage sag, whereas the latter deals with short-time duration of it, both compared with the voltage profile shown in Fig. 4 (a).

When a short-time duration fault happens, some amount of $Q$ and $P_{fault}$ must be delivered to the utility grid according to Eqs. (9) and (10), respectively. At this point, if $P_{fault}$ is greater than $P_{max}$ the MPPT algorithm is activated and all the available power of the PV generator is fed to the utility grid ($P^* = P_{max}$) setting the DC link voltage reference command to $V_{DC}*_{maximum\ power}$. On the contrary, if $P_{fault}$ is equal or less than $P_{max}$, the non-MPPT algorithm is activated instead, limiting the power delivered to the grid to $P^* = P_{fault}$ and reaching the DC link voltage reference command ($V_{DC}*_{fault}$) when the balance of power is achieved.

It is worth noting that for deep voltage sags, $Q$ can be greater than $S_{fault}$ in eq. (10) and in this case the instantaneous powers are limited to $Q = S_{fault}$ and $P_{fault} = 0$.

Figure 6 (a) shows the $P$-$V$ curve of the operation with MPPT and non-MPPT algorithms for normal and faulty operation modes, respectively. During faults, the amplitude of the PS of the three-phase utility grid voltages decreases and the outermost DC voltage regulator will try to increase the amplitude of the output inverter currents in order to do the proper power balance between the PV generator and the grid for normal operation. However, due to the current limitation imposed by the grid code, the faulty operation mode is activated assuring the correct power balance because the output voltage of the PV generator will move to the right side in the $P$-$V$ curve (from ($V_{DC}*_{maximum\ power}$, $P_{max}$) to ($V_{DC}*_{fault}$, $P_{fault}$)) increasing the DC link voltage. For deep voltage sag, $P_{fault} = 0$ and no active power is delivered to the utility grid, yielding the maximum DC voltage reference command $V_{DC}*_{max}$.

Finally, Fig. 6 (b) explains the flowchart of the LVRT deeply during the voltage sag through the analysis of a $P$-$Q$ curve, where three well-defined situations, not previous deeply studied in the scientific literature as far as the authors are concerned, are described:

1) Normal operation mode (**no voltage sag**), *Fault Signal = 0*:
   - $Q^* = 0$
   - $P^* = P_{max}$ (<u>MPPT algorithm is used</u>)
2) Faulty operation mode (**moderate voltage sag**), $Q_1 < S_{fault1}$, $P_{fault1} < P_{max}$, *Fault Signal = 1*:
   - $Q^* = Q_1$
   - $P^* = P_{fault1}$ (<u>non MPPT algorithm is used</u>)
3) Faulty operation mode (**deepest voltage sag**), $Q_2 > S_{fault2}$, *Fault Signal = 1*:
   - $Q^* = S_{fault2}$
   - $P^* = P_{fault2} = 0$ (<u>non MPPT algorithm is used</u>)

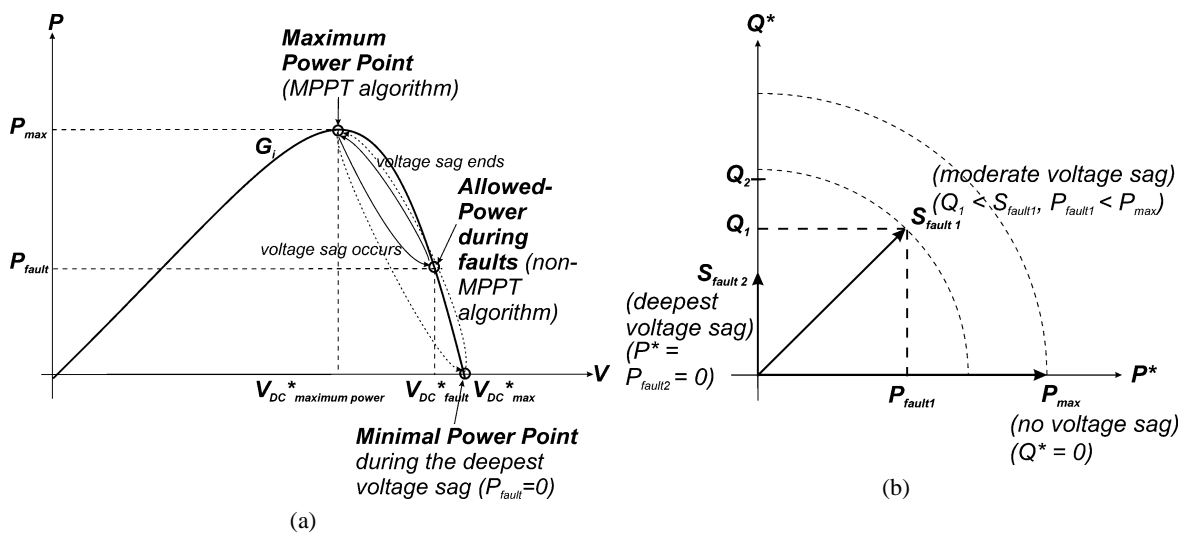

Fig. 6. (a) *P-V* curve for the grid-connected PV system, (b) *P-Q* curve of the grid-connected PV system.

## 4. Realized case study

Several PV modules and its equivalent circuits are well-described in the scientific literature [38–40], and the methods and equations used to obtain the parameters of the equivalent circuit are shown in [41,42]. From the datasheet of a **STP320-24-Ve polycrystalline** PV module manufactured by **Suntech** [43] at standard test conditions (STC) (irradiance $G = 1000\ W/m^2$ and temperature $T = 25\ °C$), the array sizing of the PV generator used in this work is calculated and shown in Table 1. A maximum DC output power of about 507 kW is attained for the DC bus voltage ($v_{DC} = 807.4\ V$) and the output current ($i_P = 627.84\ A$).

Table 1. Parameters of the PV generator.

| Parameters | Value |
|---|---|
| Number of the strings | 72 |
| Number of the series modules | 22 |
| Current at the maximum power | 627.84 A |
| Voltage at the maximum power | 807.4 V |
| DC output power | 506.91 kW |
| Open circuit voltage | 1003.2 V |
| Short circuit current | 653.04 A |

A MATLAB/SIMULINK [44] PV generator model with the specifications shown in the above Table is used to generate the several *I-V* and *P-V* curves for different irradiances and *T = 25 ℃*. These curves are used to build up a Lookup Table (2-D) in order to evaluate the proposed control algorithms under several test conditions.

Tables 2 and 3 list the parameters of the power and the control subsystems, respectively.

Table 2. Parameters of the power subsystem ($v_{DC} \approx 807\ V$ and $i_P \approx 625\ A$).

| Parameters | Value |
|---|---|
| Link capacitor *(C<sub>link</sub>)* | 65000 μF |
| Inverter Gain *(K<sub>PWM</sub>)* | $\frac{2}{3}\text{v}_{\text{DC}}$ |
| Switching Frequency *(f<sub>sw</sub>)* | 24.416 kHz |
| Line Inductance *(L)* | 0.15 mH |
| AC system *(u<sub>grst</sub>)* | -   Voltage level: 230 V(*rms*) phase-to-neutral<br>-   Frequency: 50 Hz and 60 Hz |

Table 3. Parameters of the control subsystem ($f_{CI} = 610.4\ Hz$, $f_{CV} = 12.208\ Hz$, $PM_I = PM_V = 63.5°$).

| Parameters | Value |
|---|---|
| Proportional constant of the PR controller *(K<sub>Paβ</sub>)* | 0.0011 |
| Integral constant of the PR controller *(K<sub>Iαβ</sub>)* | 0.1 |
| Resonant angular frequency of the PR controller *(ω<sub>b</sub>′)* | 314.16/377 rad/s |
| Cut-off frequency of the PR controller *(ω<sub>c</sub>)* | 1 rad/s |
| Proportional constant of the DC voltage regulator *(K<sub>PVDC</sub>)* | 3977.5 |
| Integral constant of the DC voltage regulator *(K<sub>IVDC</sub>)* | 152110 |
| Gain of the SOGI-QSG blocks *(K/i)* | $\sqrt{2}$ |
| Gain of the normalization block inside the FLL *(K)* | $\sqrt{2}$ |
| Gain of the FLL *(Γ)* | 50 |
| Sample time of the power subsystem *(T<sub>S</sub>)* | 5.1196 μs |
| Sample time of the control subsystem *(T<sub>reg</sub>)* | 40.9568 μs |

## 5. Simulations

In this section, some simulations are done in order to validate the control strategy employed.

Figures 7 and 8 depict the behavior of the single-stage grid-connected PV system during a voltage sag of 50% of amplitude in phase 3 for an irradiance $G_1 = 1000\ W/m^2$ and $G_2 = 800\ W/m^2$, respectively. The LVRT capability is validated in both situations where a constant active power operation and oscillating components at $2\omega_0´$ in $Q$ are achieved, although the three-phase grid currents delivered to the utility grid are unbalanced but with a limitation in their amplitudes to the nominal value. At the same time, the constant active power operation will also reduce the harmonic distortions at $2\omega_0´$ in the DC link voltage and the DC current from the PV generator.

Figure 9 depicts the behavior of the system when a distortion of 10% in the amplitude of the 5th and the 7th harmonics of the three-phase grid voltages happen, achieving undistorted three-phase grid currents, constant active power and oscillations at $6\omega_0´$ in $Q$.

Finally, Fig. 10 depicts the behavior when a variation of the incoming irradiance from $1000$ to $800\ W/m^2$ happens (i.e, a shadow). $P$ varies from 500 kW before the step of the irradiance to 400 kW after the step, keeping $Q = 0$.

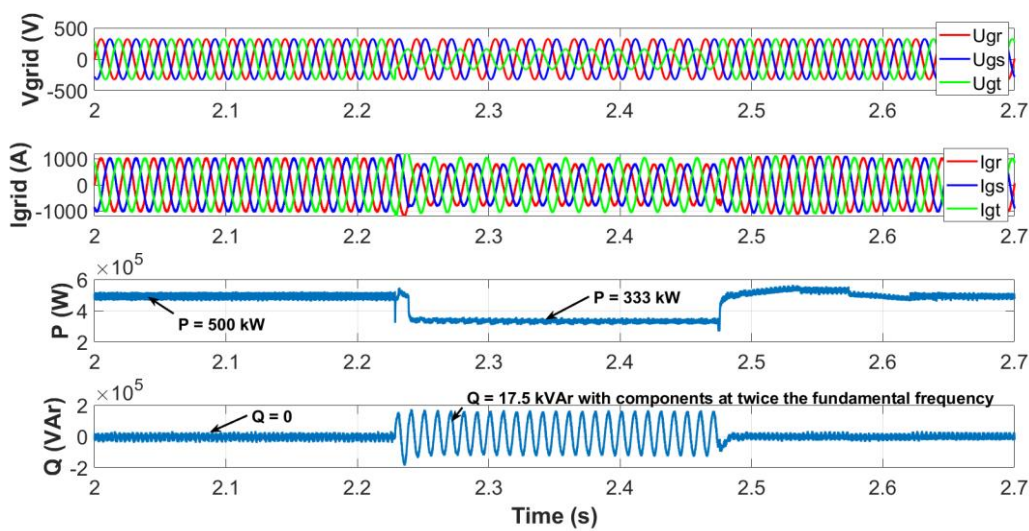

Fig. 7. Output variables during a voltage sag of 50% of amplitude in phase 3 ($G_1 = 1000\ W/m^2$).

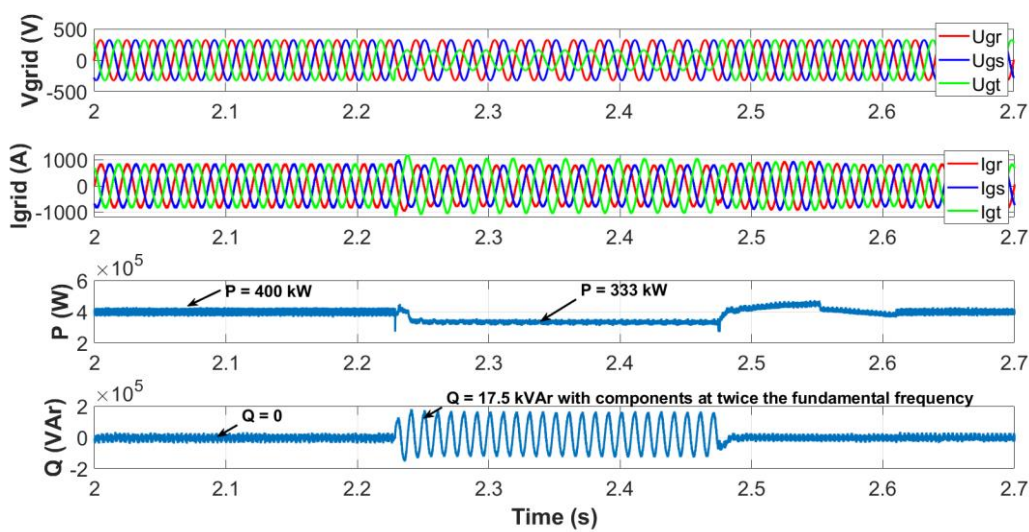

Fig. 8. Output variables during a voltage sag of 50% of amplitude in phase 3 ($G_2 = 800\ W/m^2$).

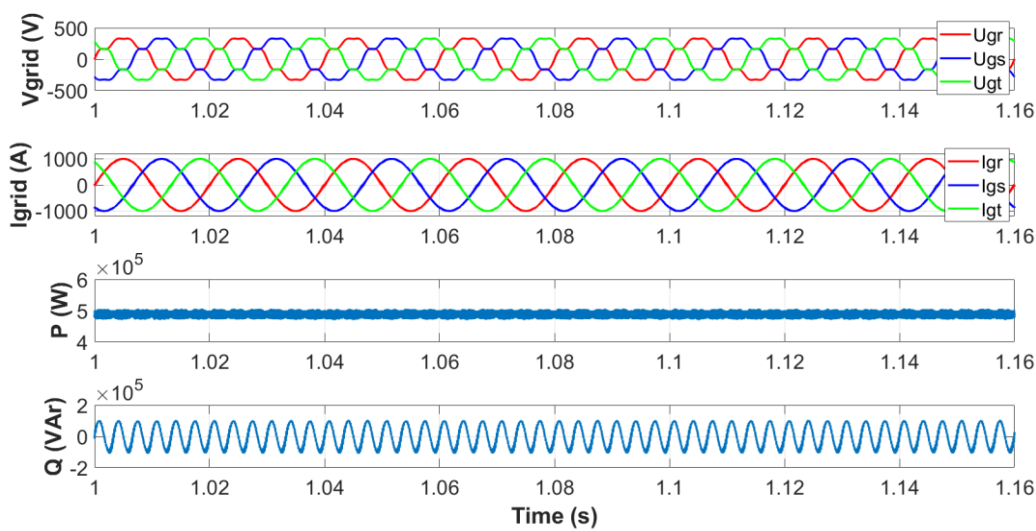

Fig. 9. Output variables for a 10% amplitude in the 5th and 7th harmonics in the utility grid.

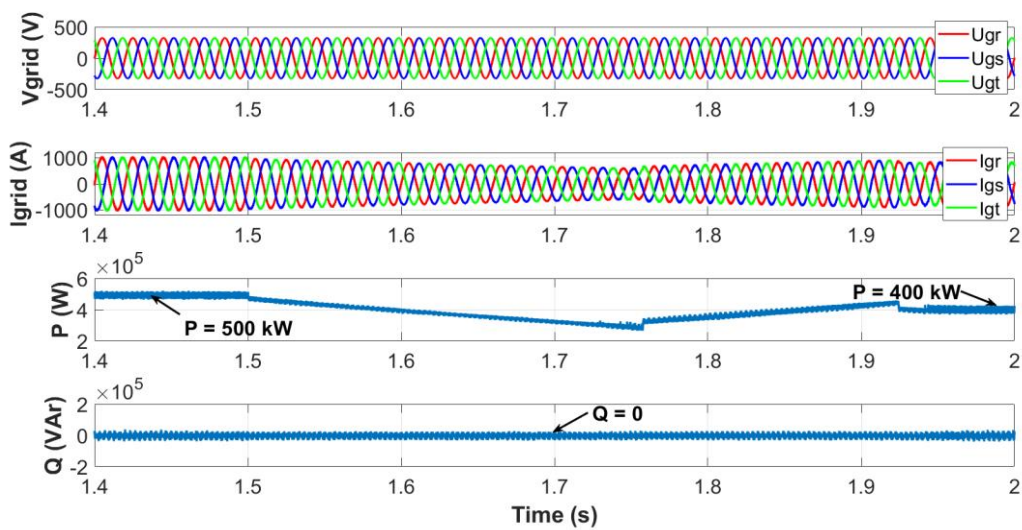

Fig. 10. Output variables for a step in irradiance G (shadow) from 1000 to 800 kW/m².

## 6. Experiments using the CHIL technique

Some tests will be carried out in this Section with the **Controller-Hardware-in-the-Loop (CHIL)** simulation technique [10,45] in which the power subsystem is modelled with PLECS tools and the control subsystem is modelled using Simulink blocks. C-code is generated from both models and downloaded into the PLECS RT Box 1 [46] and into the TMS320F28379D Dual-Core Delfino Microcontroller [47]. The former behaves as the Digital Real-Time Simulator (DRTS) [48–50], whereas the latter works as the real microcontroller, running both in real-time [51].

This scenario is more realistic if compared with functional simulations because the influence of some second order issues such as sample delays, dead-time of the inverter, quantization errors of the ADCs, etc., not modelled in simulations, can be investigated. On the other hand, the safety of the platform and people are assured, as well as the debugging of complex and risky control algorithms before the final prototype were built.

It is worth noting that an ideal model for the three-phase VSI has been used in the implementation of the CHIL technique in this paper, which does affect the results for the proper validation of the control algorithms. The estimated switching power losses for the SiC MOSFET-based *SKM350MB120SCH17* module employed [52] is about 2436 W when switching at 24.416 kHz, allowing an efficiency higher than 99%.

The block diagram of the CHIL setup is shown in Fig. 11, where 9 analog signals with the real-time behavior of several voltages and currents in the DRTS are sent from its analog outputs to the microcontroller. Then, 3 digital signals with the status of the power switches are sent from the pulse width modulation (PWM) peripheral of the microcontroller to the DRTS, closing the control loops.

The complexity of the model used for the power subsystem and the limitations of the hardware resources of the used DRTS impose a sample time of at least 10 μs. So, the sample time is set to 5.1196 μs for the DRTS ($T_s$) and to 40.9568 μs for the controller ($T_{reg}$) which is eight times greater. The inverse of $T_{reg}$ has also the same value as the switching frequency ($f_{sw}$) allowing the synchronization between the sample time of the currents and the digital PWM and avoiding the use of low-pass filters to the sampled currents; the controller has a delay of $\frac{3}{2} T_{reg} = 61,44 \ \mu s$ in the worst case [53]. Special care must be considered for the discretization of the control algorithms, and they must use the same sample time of the microcontroller so as to be able to avoid the reduction of the $PF$. Moreover, the measuring errors of the electronic card can introduce unbalances [54]. Finally, a photography of the CHIL platform is shown in Fig. 12.

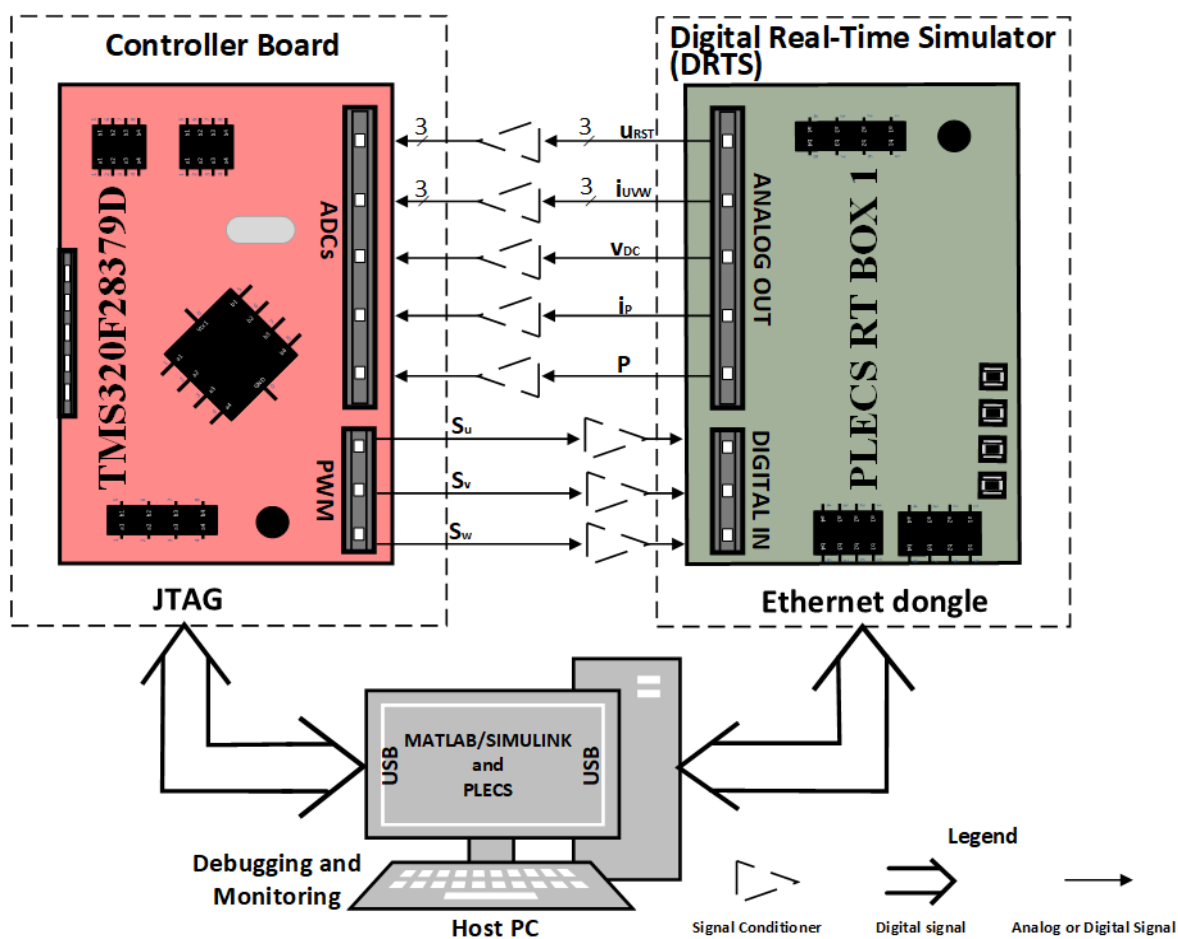

Fig. 11. Block diagram of the CHIL simulation setup.

Several measurements in the output signals of the power subsystem are carried out, displayed and recorded in the host PC for monitoring purposes, at STC ($G = 1000 \ W/m^2$ and $T = 25 \ ^\circ C$) in the PV generator, although the measurements can be made for any other irradiance. Some different types of voltage sags are exerted in order to validate the LVRT capability: a three-phase deep voltage sag of 90% in its amplitude, a deep voltage sag in phase 3 of 90% in its amplitude, and a moderate voltage sag in phase 3 of 50% in its amplitude.

Figures 13-15 depict the behavior of the grid-connected inverter when the utility grid voltages have a nominal frequency of 50 Hz, whereas Figs. 16-18 depict the behavior for a nominal frequency of 60 Hz.

The evolution in time of the DC link voltage and current at the output of the PV generator, the three-phase grid voltages and currents, and $P$ and $Q$ delivered to the grid when a three-phase deep voltage sag of 90% in amplitude happens are depicted in Figs. 13 and 16. During the voltage sag operation mode the three-phase currents lag the grid voltages by 90° in both figures and the $PF$ of the system is zero in this situation. So, no active power is delivered to the utility grid, whereas $Q = 50 \ kVAr$ is delivered to the grid in order to improve the voltage profile, as shown in both figures. On the contrary, in the normal operation mode the three-phase grid voltage and currents are synchronized yielding a connection with an unity *PF* (*Q = 0*) and with the maximum *P* delivered to the grid.

Figures 14 and 17 depict the same variables in (a) when a deep voltage sag of 90% in amplitude happens in phase 3, whereas the time evolution of these variables during the voltage sag are depicted in (b). Moreover, Figs. 15 and 18 depict the same situation when a moderate voltage sag of 50% in amplitude happens in phase 3.

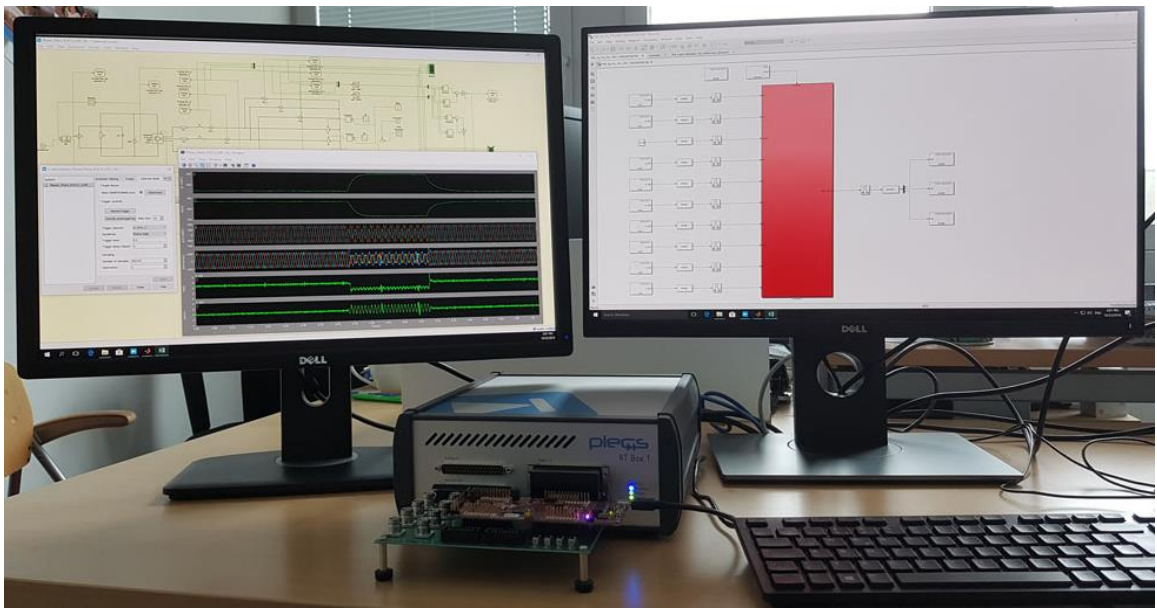

Fig. 12. Photography of the CHIL platform.

In the normal operation mode the maximum power of the PV generator is tracked and the $V_{DC}*_{maximum\ power}$ DC link voltage command is applied to the control algorithm to do the power balance, reaching a *P* around 500 kW whereas a zero *Q* is attained for unity *PF* operation. On the other hand, the LVRT capability is activated during the voltage sags operation mode because the limitation in the amplitude of the line currents decreases the *P* injected to the utility grid and an increment in the DC link voltage is achieved because the $V_{DC}*_{fault}$ DC link voltage command is set in this situation. Some amount of *Q* is needed as well according to eq. (9) to improve the profile of the utility grid voltages during the sag.

### 6.1. Nominal frequency of 50 Hz for the grid voltages

In Fig. 13 there is no NS component of the grid voltages due to the three-phase voltage sag (also known as balanced fault) and then, *P* and *Q* have only constant components. However, in Figs. 14 and 15, the voltage sag in phase 3 (known as unbalanced fault) produce oscillations not only in *P* and/or *Q*, but also in the DC link voltage ($v_{DC}$) and in the DC current from the PV generator ($i_{DC}$). The oscillations are at twice the system frequency ($2\omega_0´\hookrightarrow$period *T = 10 ms*, $\omega_0´ = 2\pi 50 = 314.16\ rad/s$) of the grid voltages due to its NS component under unbalanced faults and are added to the average constant values of the several variables involved. In this paper, a constant active power operation is chosen according to eq. (5) as said before, producing strong oscillations at $2\omega_0´$ in *Q* around its average value, whereas light oscillations at $2\omega_0´$ appear in *P*, $v_{DC}$ and $i_{DC}$. These light oscillations are due to the small filter´s resistance of the inductive filter that consumed small oscillations delivered by the inverter [21].

The constant active power operation also produces unbalanced sinusoidal three-phase grid currents when unbalanced fault happen, as can be seen in Figs. 14 and 15, whose amplitudes are limited to its nominal value to avoid the triggering of the overcurrent protection of the inverter and its disconnection from the mains during the voltage sags, as it was analyzed before.

It can also be noticed the increase of the DC link voltage for both the balanced and unbalanced faults, according to the effect of the non-MPP algorithm during the voltage sags, as it is described in Fig. 6 (a).

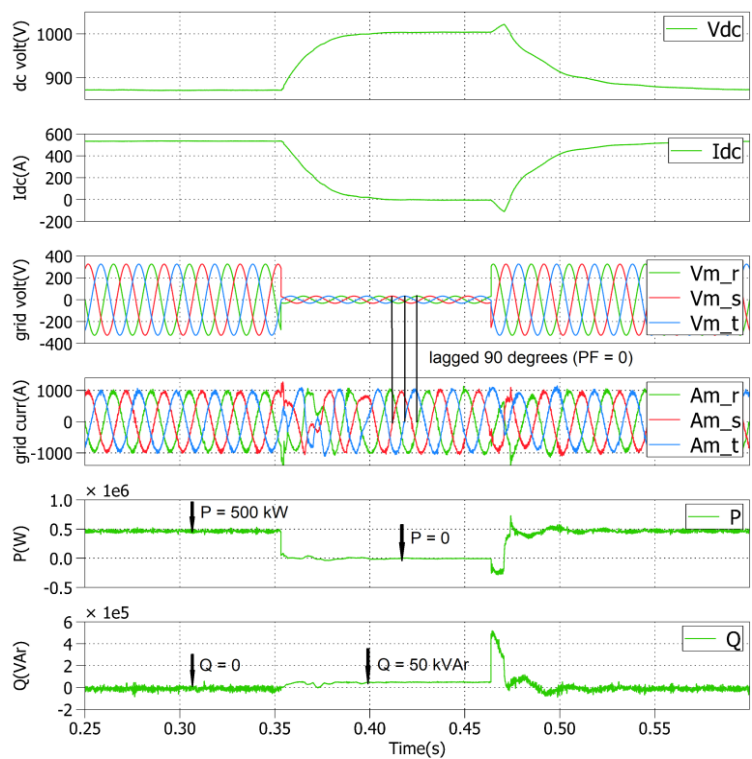

Fig. 13. Evolution of the voltages, currents and powers during a three-phase deep voltage sag of 90% in amplitude.

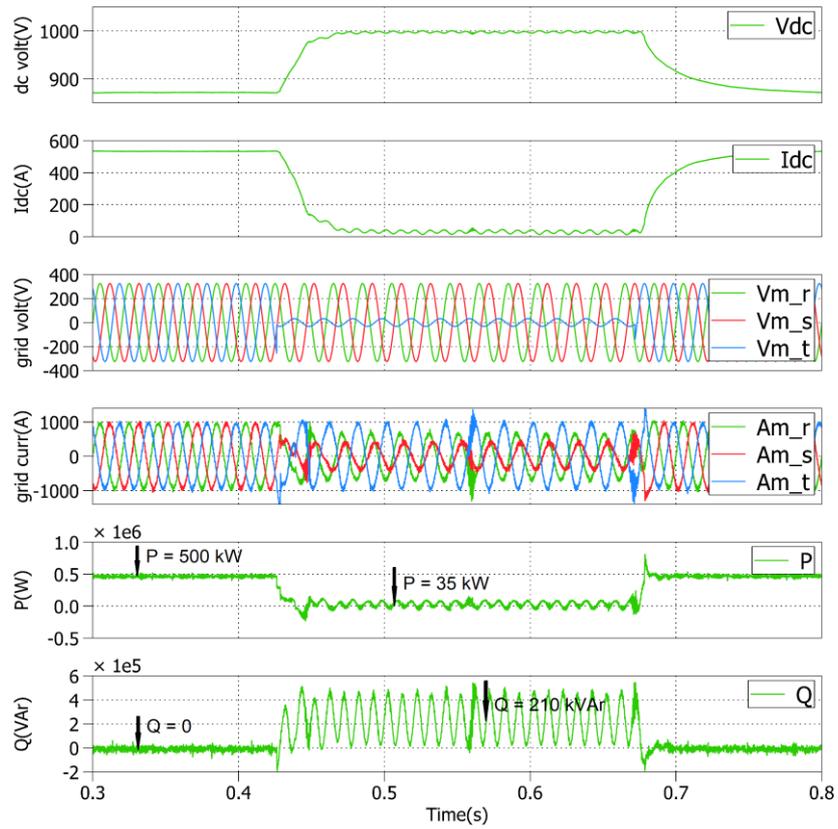

(a)

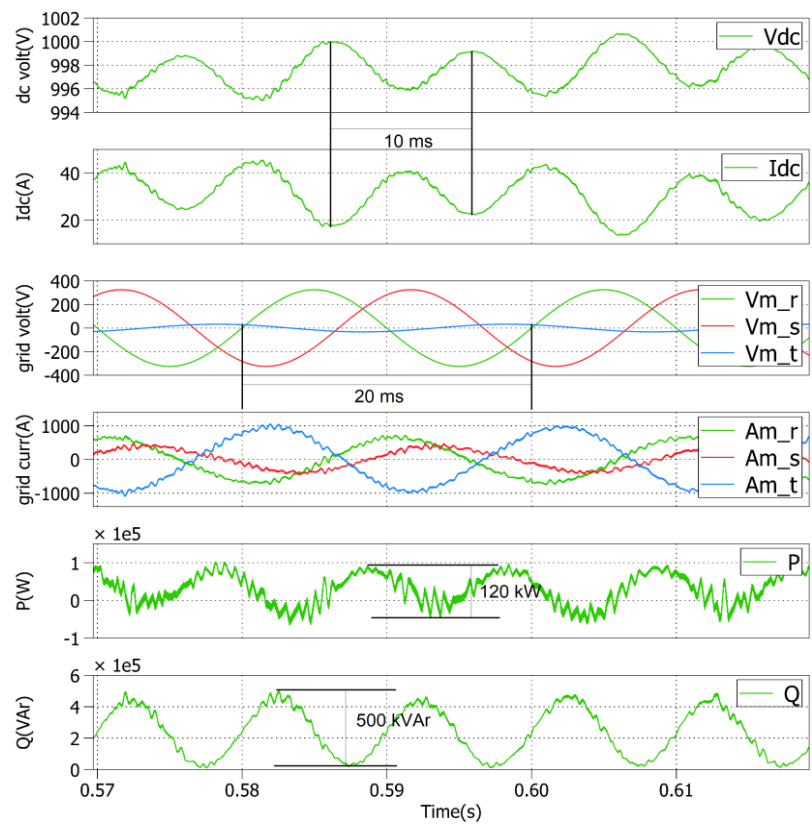

(b)

Fig. 14. (a) Evolution of the voltages, currents and powers during a deep voltage sag in phase 3 of 90% in amplitude, (b) Evolution of the same variables during the voltage sag.

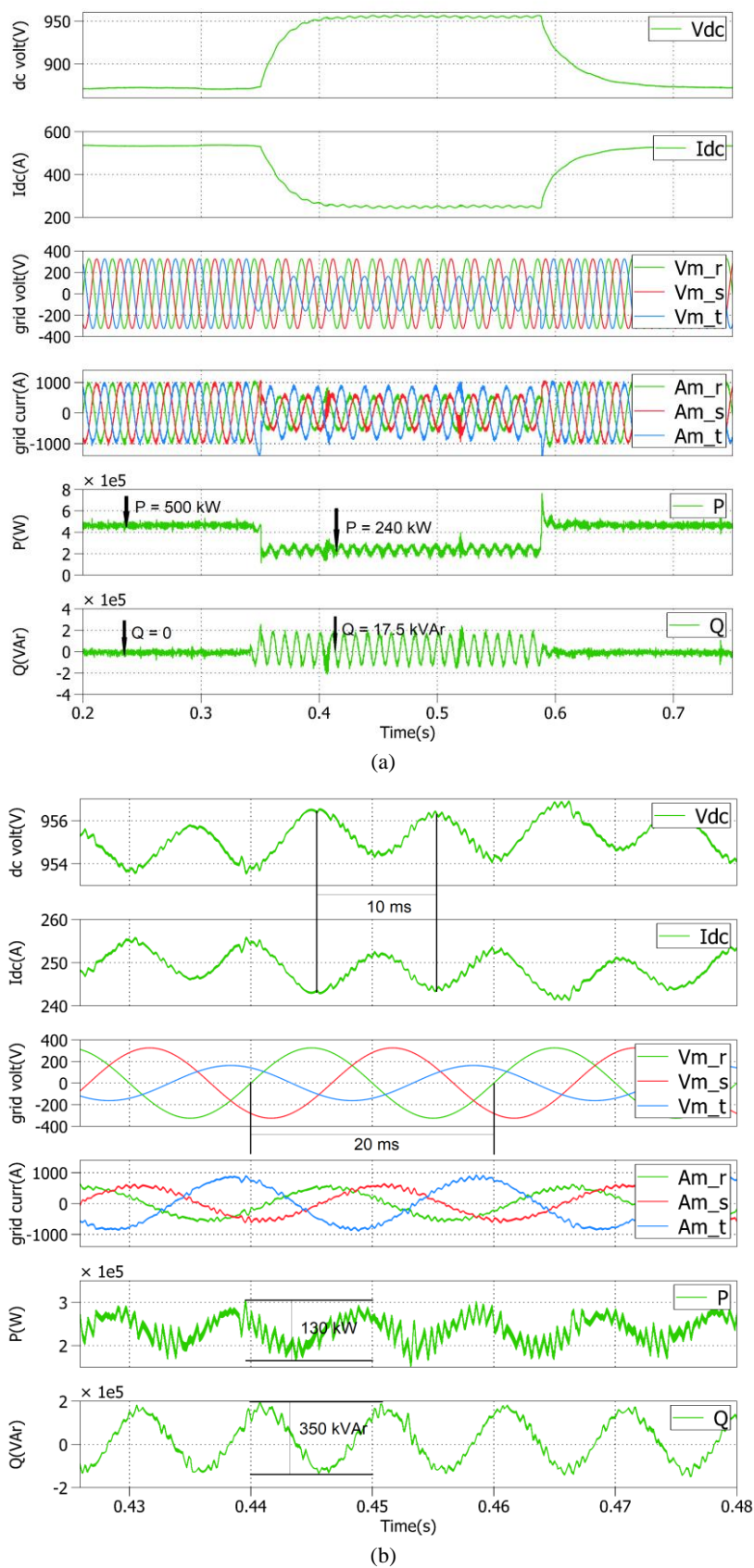

(a)

(b)

Fig. 15. (a) Evolution of the voltages, currents and powers during a moderate voltage sag in phase 3 of 50% in amplitude, (b) Evolution of the same variables during the voltage sag.

*6.2. Nominal frequency of 60 Hz for the grid voltages*

The same explanation for the nominal frequency of 50 Hz made in Section 6.1 is valid here. In this case the period $T = 8.33\ ms$ for $2\omega_0'$, where $\omega_0' = 2\pi \cdot 60 = 376.992\ rad/s$.

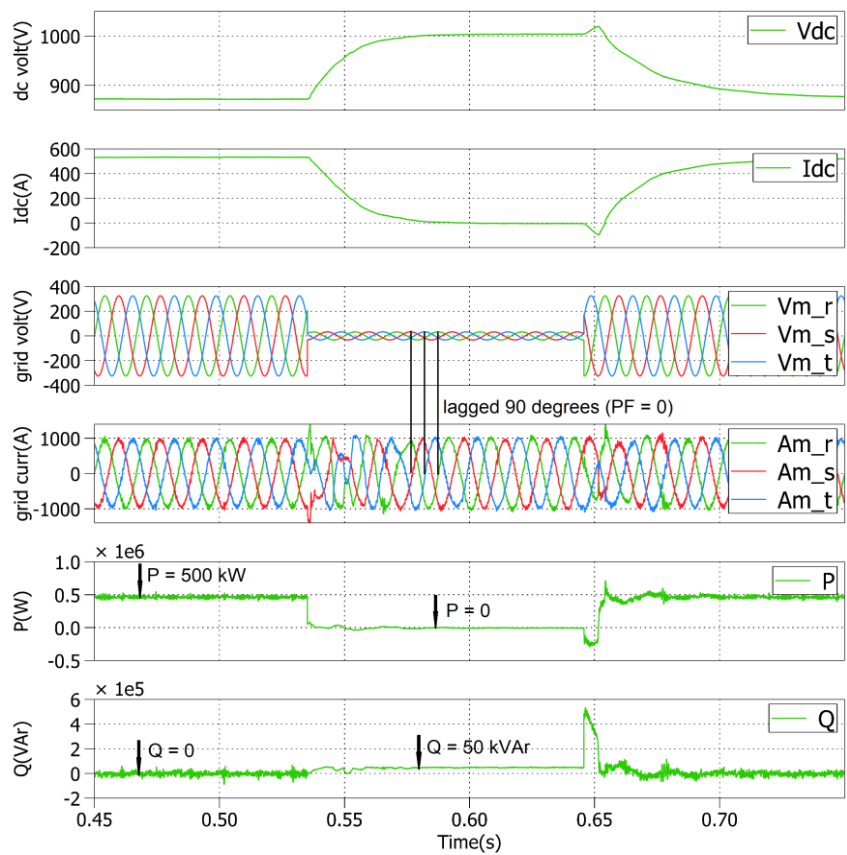

Fig. 16. Evolution of the voltages, currents and powers during a three-phase deep voltage sag of 90% in amplitude.

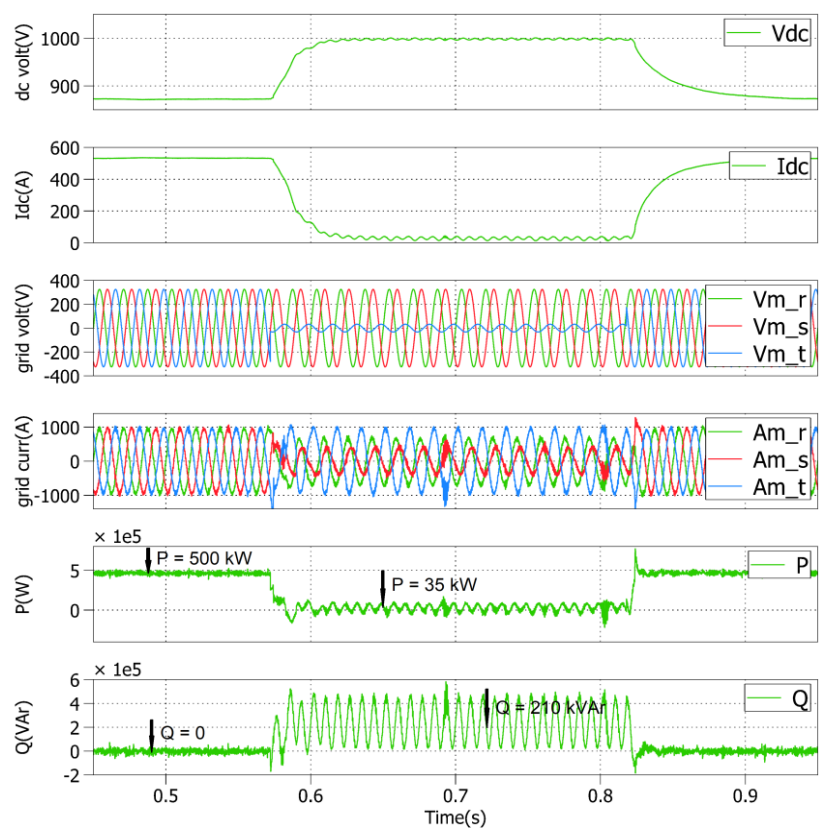

(a)

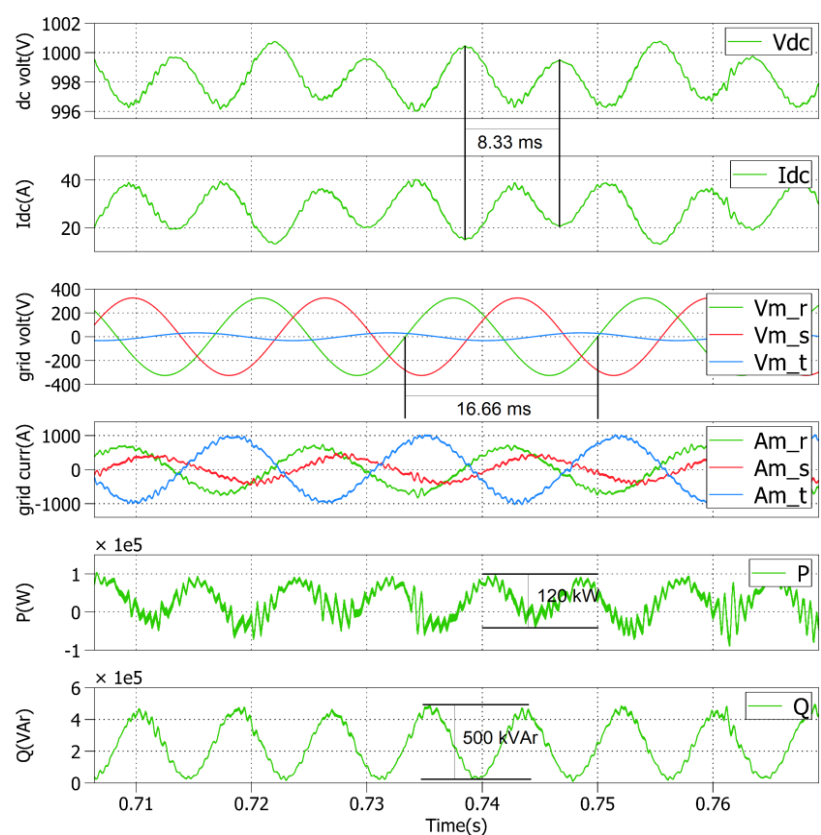

(b)

Fig. 17. (a) Evolution of the voltages, currents and powers during a deep voltage sag in phase 3 of 90% in amplitude, (b) Evolution of the same variables during the voltage sag.

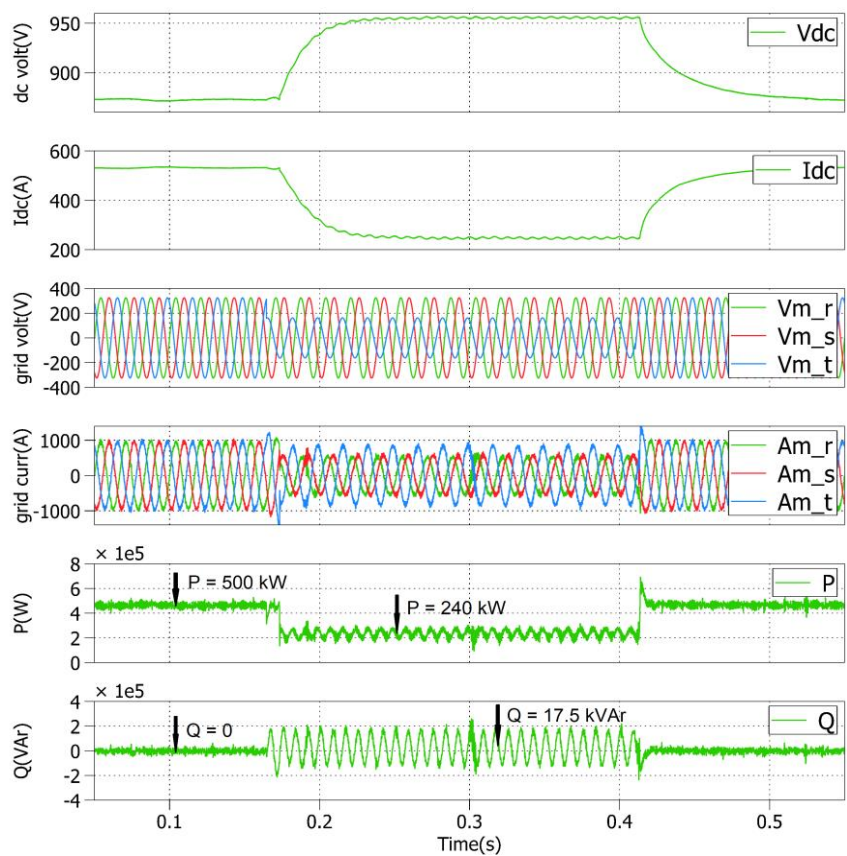

(a)

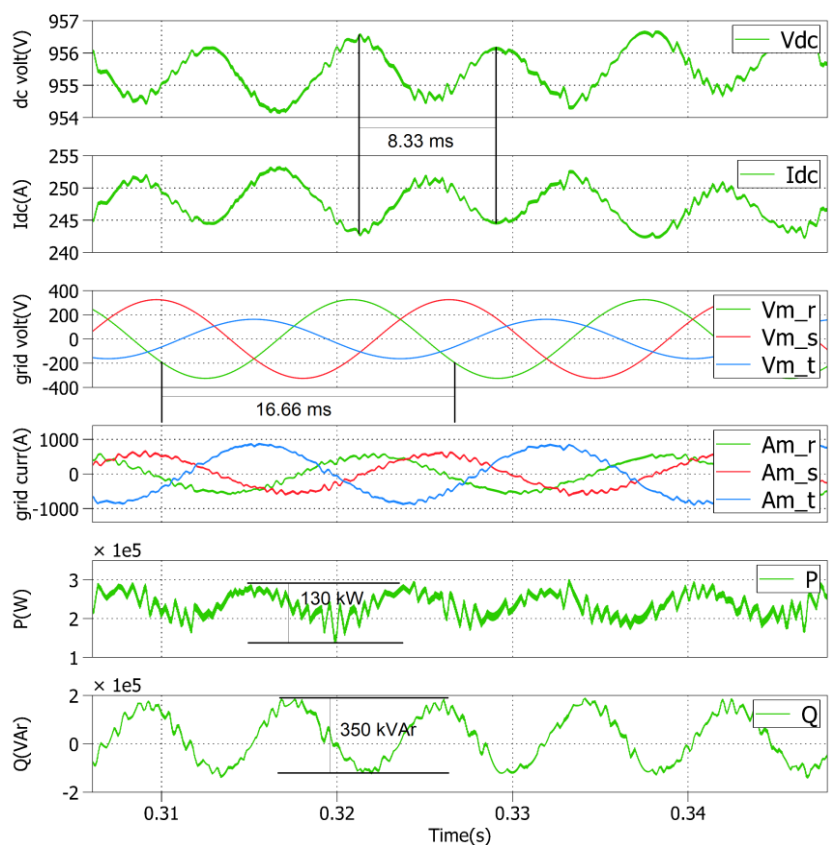

(b)

Fig. 18. (a) Evolution of the voltages, currents and powers during a moderate voltage sag in phase 3 of 50% in amplitude, (b) Evolution of the same variables during the voltage sag.

## 7. Conclusions

In this article the LVRT capability was tested as an ancillary service for the control algorithm of a three-phase single-stage grid-connected PV system under grid voltage sags. This service was carried out according to the IEC 61400-21 European normative and the spanish grid code during the faulty operation mode, although the methodology employed can be easily applied to any other international normative. For this, the detection of the PS and the NS components of the three-phase grid voltages is a most in order to properly calculate the PS and the NS of the three-phase inverter currents. Hence, a constant active power control can be exerted to protect the DC link capacitor from large oscillations at twice the system frequency which could destroy it. In addition, a certain amount of reactive power must also be injected into the utility grid to facilitate the improvement of the voltage profile in a Distributed Generation scenario.

The MSOGI-FLL synchronization algorithm used in this paper computes the PS and the NS components of the three-phase grid voltages in $\alpha\beta$ axes, as well as the system frequency, both free of harmonic distortions. The estimated frequency is fed back to the PR current regulators and to the HC structure, making the system frequency-adaptive.

The limitation in the amplitude of the three-phase inverter currents for both the faulty and non-faulty operation modes has been studied in the paper and four well-defined situations arise: 1) no voltage sags, 2) three-phase deep and moderate voltage sags, 3) three-phase small voltage sags, 4) unbalanced voltage sags. For the first three cases a maximum nominal value for the amplitude of the reference currents is achieved, which describes a *circumference* in $\alpha\beta$ axes, whereas for the last case an *ellipse* with less amplitudes than the maximum nominal value seen in the previous situations is attained, guaranteeing the limitation in the amplitude of the three-phase inverter currents after the application of the inverse Clarke's transformation. The results of these four cases can be extended to any grid-connected PV system under unbalanced grid voltage sags in order to exert, for example, the proper control to deliver constant and/or oscillations $P$ and $Q$, if it were necessary.

A comprehensive flowchart for the LVRT ancillary service is shown in the paper and its main results can be deduced from the *P-Q* curve for three situations: 1) normal operation mode for no voltage sag in which $Q^* = 0$ and the MPPT algorithm is used; 2) faulty operation mode for moderate voltage sag in which $P^* = P_{fault}$, a certain amount of reactive power is delivered to the utility grid ($Q^* \neq 0$), and the non MPPT algorithm is used; 3) faulty operation mode for the deepest voltage sag, in which $P^* = 0$, a small value of reactive power is delivered to the utility grid ($Q^* \neq 0$) and the non MPPT algorithm is also used.

Finally, some simulations are doing with MATLAB/SIMULINK and some experiments are carried out with a realistic validation setup using CHIL simulations for a realized case study of a 500 kVA three-phase single-stage grid-connected PV system, in which the theoretical findings are validated.

### Acknowledgments

This work was supported by the project "Nuevas topologías para convertidores en MT para grandes Instalaciones Fotovoltaicas" from the Spanish Government (Ref. TEC2016-80136-P) (A. B. Rey-Boué) and the European Community's Horizon 2020 Program (H2020/2014-2020) in project "ERIGrid" (Grant Agreement No. 654113) under the Trans-national Access (TA) User Project: 04.003-2018.